\documentclass[aip,rsi,reprint,graphicx]{revtex4-1}
\usepackage[colorlinks=true,linkcolor=blue]{hyperref}%
\usepackage{graphicx}

\newcommand{\textss}[1]{\scriptsize \mbox{#1}}
\newcommand{\YuningPulse}{85}
\newcommand{\XinlongPulse}{155}

\begin{document}

\title{High-power ultrafast Yb:fiber laser frequency combs using commercially available components and basic fiber tools}

\author{X. L. Li, M. A. R. Reber, C. Corder, Y. Chen, P. Zhao, and T. K. Allison}

\email{thomas.allison@stonybrook.edu} 
\affiliation{Stony Brook University, Stony Brook, NY 11794-3400 USA}

\date{\today}%

\begin{abstract}

We present a detailed description of the design, construction, and performance of high-power ultrafast Yb:fiber laser frequency combs in operation in our laboratory. We discuss two such laser systems: an 87 MHz, 9 W, $\YuningPulse$  fs laser operating at 1060 nm and an 87 MHz, 80 W, \XinlongPulse fs laser operating at 1035 nm. Both are constructed using low-cost, commercially available components, and can be assembled using only basic tools for cleaving and splicing single-mode fibers. We describe practical methods for achieving and characterizing low-noise single-pulse operation and long-term stability from Yb:fiber oscillators based on nonlinear polarization evolution. Stabilization of the combs using a variety of transducers, including a new method for tuning the carrier-envelope offset frequency, is discussed. High average power is achieved through chirped-pulse amplification in simple fiber amplifiers based on double-clad photonic crystal fibers. We describe the use of these combs in several applications, including ultrasensitive femtosecond time-resolved spectroscopy and cavity-enhanced high-order harmonic generation.
 
\end{abstract}

\maketitle

\section{Introduction}

Originally intended for the precise measurement of optical frequencies, femtosecond optical frequency combs have since found many other applications outside of their original purpose.\cite{Newbury_NatPhot2011,Diddams_JOSAB2010} They are now used for the calibration of astronomical spectrographs,\cite{Steinmetz_Science2008} laser ranging,\cite{Coddington_NatPhot2009} high-order harmonic generation,\cite{Mills_JPhysB2012} attosecond physics,\cite{Krausz_RMP2009, Benko_NatPhot2014} and direct frequency comb spectroscopy,\cite{Adler_AnnRevChem2010, Coddington_Optica2016} among other things. In our lab, for example, we have recently demonstrated a large improvement in the sensitivity of ultrafast optical spectroscopy using frequency comb methods.\cite{Reber_Optica2016} Most frequency comb lasers operate in the near infrared, based on Ti:sapphire, Er:fiber, or Yb-based gain media, but most spectroscopic applications of frequency combs lie in other regions of the electromagnetic spectrum. It is then desirable to shift the comb to other spectral regions using nonlinear optical techniques,\cite{Adler_OptLett2009, Cingoz_Nature2012, LeinDecker_OptExp2012, Cruz_OptExp2015} but doing this with high efficiency requires high peak powers. Thus for a frequency comb with a useful repetition rate and comb spacing, high average power is needed.    

Ytterbium-based systems then stand out as providing an excellent platform for average power scaling due to the very small quantum defect of Yb and the capability of high doping in both glasses and crystals. \cite{Paschotta_IEEE1997, Fermann_IEEE2009, Fattahi_Optica2014} Since the first Yb-based femtosecond lasers,\cite{Cautaerts_OptLett1997, Honninger_OptLett1995} progress in this field has moved at a rapid pace using fibers, thin disks, and slabs. Yb:fibers are particularly attractive for average power scaling, due to the large surface area to volume ratio of fibers, large gain bandwidth,\cite{Paschotta_IEEE1997} and the availability of double-clad fibers for use with low-brightness (and thus low-cost) pump diodes. Indeed, kW scale femtosecond lasers have been reported using high-power Yb:fiber amplifier systems,\cite{Eidam_OptLett2010,Jauregui_nph2013} and amplified Yb-combs have demonstrated high phase coherence.\cite{Schibli_NatPhot2008} Yb:fiber has sufficient gain bandwidth to support sub-200 fs pulses through linear chirped pulse amplification (CPA), in which the total nonlinear phase shift accumulated in the amplifier chain is kept less than 1 radian.\cite{Kuznetsova_APB2007} Recently, shorter pulses have been generated at high average power through nonlinear amplification,\cite{Zhao_OptExp2014, Liu_OptLett2015} in which nonlinear propagation in the gain fibers is harnessed for generating additional spectral components. With narrow linewidth and controllable combs,\cite{Schibli_NatPhot2008, Hartl_OptLett2007, Ruehl_OptLett2010, Nugent-Glandorf_OptLett2011} the simultaneous combination of high peak and average power can be obtained through enhancement in passive optical  cavities.\cite{Jones_OptLett2002, Jones_OptLett2004, Carstens_OptLett2014,Hartl_OptLett2007} 

The literature regarding this development has been confined to conference proceedings and specialty optics journals, and it requires quite a bit of know-how to go from this literature to a working femtosecond fiber laser. Indeed, most femtosecond lasers used in laboratory research are still based on Ti:sapphire, and there is a much larger community that is familiar with the operation of Ti:sapphire lasers than fiber-based systems. One can build a  fiber laser for a fraction of the cost of an equivalent Ti:sapphire system,\cite{Paul_OptLett2008} and it is much simpler, but there is a learning curve to climb: What pump diodes do I use? Is that cheap current controller quiet enough? How do you mount the fibers? How fast does the pump diode need to turn off to save the fiber amplifier if the seed is lost? These are the types of questions we encountered when starting to build fiber lasers in 2013 - and they can be fiendishly difficult to find in the literature. 

\begin{figure}[t]
\centering
\includegraphics[width=\linewidth]{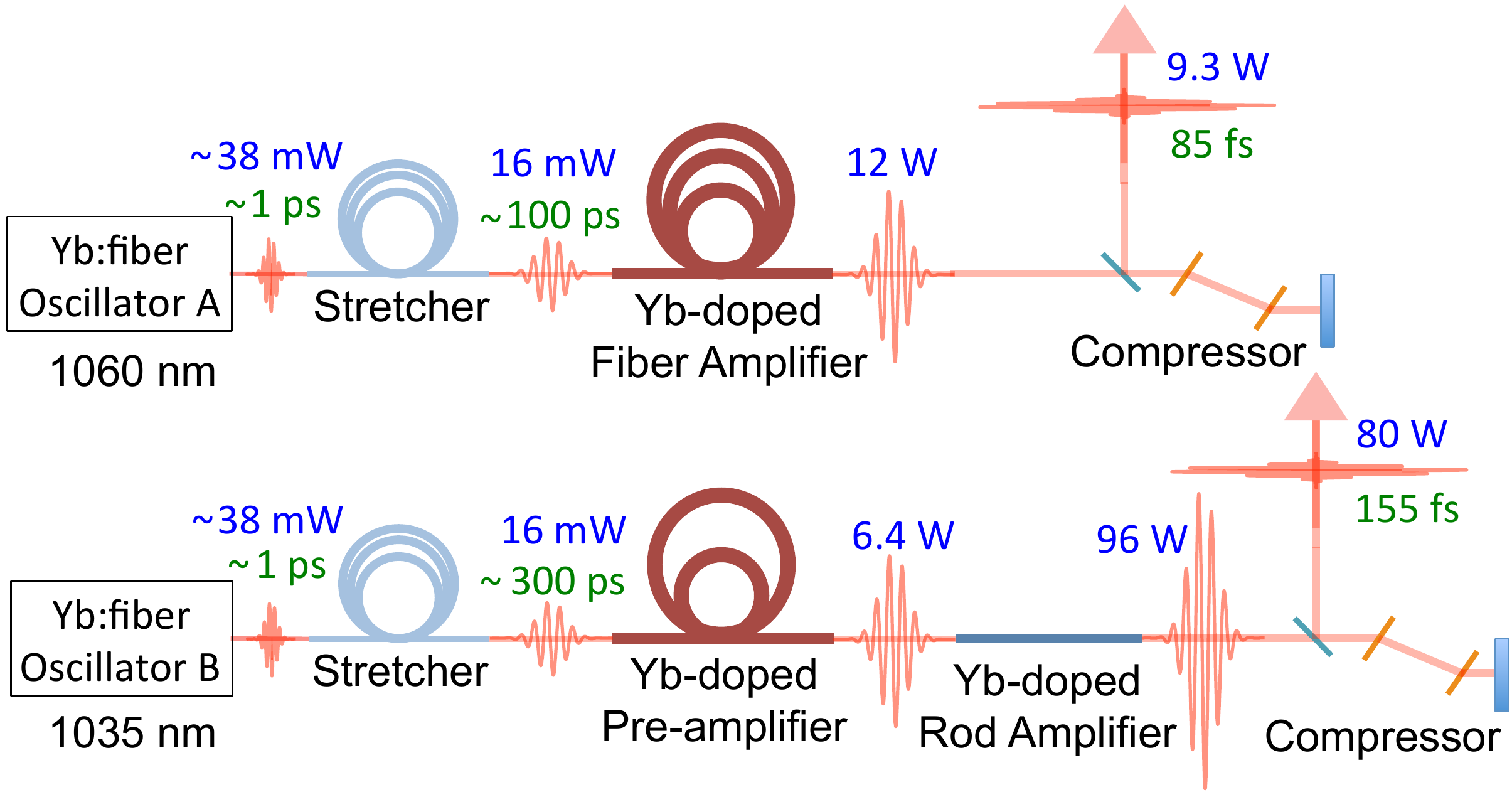}
    \caption{\small \textbf{Overview of the laser systems.} Chirped pulse amplification using a fiber-based stretcher and simple PCF amplifiers provides minimal complexity. More detailed schematics are shown in figures \ref{fig:osc}, \ref{fig:10Wamp}, and \ref{fig:80Wamp} and component lists and mechanical drawings are provided  in the supplemental material.}
    \label{fig:overview}
\end{figure}

In this article we present a detailed account of the design, construction, and operation of two high-power Yb:fiber laser frequency combs that we built in our laboratory over the past three years. The literature regarding Yb:fiber lasers is vast, and we are not attempting here to provide a comprehensive review. Rather, we hope to provide a practical guide to those not already intimately familiar with the details of fiber lasers. A block-diagram outlining both lasers is shown in figure \ref{fig:overview}. The lasers we describe in this article are made from all commercial components, most of which are stock items at major distributors, and can be assembled with only basic fiber tools. For example, a fiber splicer capable of splicing standard single-mode fibers is sufficient and you do not need more expensive models capable of handling polarization maintaining fibers or photonic crystal fibers (PCF), which cost many times more.

In section \ref{sec:osc}, we describe the construction of Yb:fiber oscillators passively  mode locked using nonlinear polarization evolution (NPE) and run near zero net cavity group delay dispersion (GDD) for the lowest noise and narrowest comb-tooth linewidth.\cite{Nugent-Glandorf_OptLett2011} The physics and many implementations of these oscillators has been discussed extensively in the literature (for a review see reference \citenum{Chong_RepProgPhys2015}), but NPE mode locking has a reputation for being finicky, and this literature is not necessarily helpful when standing over the laser and trying to make it work. We instead focus on practical procedures for achieving and characterizing mode-locked single-pulse operation, low noise, and long-term stability. We also discuss stabilization of the comb, including a new method for actuating on the comb's carrier-envelope offset frequency using the intracavity grating separation. In section \ref{sec:10Wamp}, we describe a simple scheme for amplifying the comb to 9 W of average power and \YuningPulse fs pulse duration at 1060 nm using CPA in a one-stage PCF amplifier. In section \ref{sec:80Wamp}, we describe CPA to 80 W using a PCF rod amplifier. Both amplifiers use fiber-based pulse stretching with anomalous third order dispersion fibers,\cite{Fernandez_OptLett2012,Ruehl_OptLett2010} which require careful initial design, but then dramatically simplify the mechanical design of the laser system. In section \ref{sec:apps}, we describe several applications of these lasers. Detailed lists of all components appear in the supplemental material.

\section{Yb:fiber oscillators}\label{sec:osc}

The large nonlinear phase shifts accumulated when an ultrashort pulse propagates for a distance in the confined space of an optical fiber can give rise to many phenomena.\cite{Dudley_NatPhys2007, Fermann_PRL2000, Agrawal_AppNonlinearFiberOptics,Agrawal_NonlinearFiberOpticsBook} Inside a laser cavity, this high nonlinearity can allow for mode-locked operation over a very wide range of parameters. Unlike Ti:sapphire oscillators, in which a few standard designs that emerged in the 1990's \cite{Asaki_OptLett1993, Stingl_OptLett1995} are found in most ultrafast laser labs, there are many fiber oscillator  designs working at repetition rates from 100 kHZ \cite{Kobtsev_OptExp2008} to 10 GHz,\cite{Usechak_OptLett2004} and the literature presents a large and daunting landscape to navigate. We attempt a brief summary here with the goal of putting our lasers in context.

Mode-locked fiber lasers can be broadly classified by their net cavity GDD and the saturable loss mechanism by which they are mode locked.\cite{Fermann_IEEE2009, Chong_OptExp2006} With large anomalous GDD, soliton-like pulse shaping produces nearly chirp-free pulses, but with limited power.\cite{Kelly_ElecLetters1992, Fermann_IEEE2009} Lasers working with large normal GDD, even with all normal dispersion elements,\cite{Chong_OptExp2006} can support wave-breaking free pulses of very large energy. For example, Baumgartl \emph{et al.}\cite{BaumGartl_OptLett2012} have even demonstrated  66 W of average power and $\mu$J pulses directly from an oscillator without subsequent amplification. However, for the quietest  operation, with both the lowest phase and amplitude noise most suitable for comb applications, it is desirable to operate the laser near net zero cavity GDD. \cite{Nugent-Glandorf_OptLett2011, Paschotta_ApplPhysB2004_1,Paschotta_ApplPhysB2004_2,Hartl_CLEO2005,Schibli_NatPhot2008} Unlike Er-doped fiber lasers operating at 1.5 $\mu$m, where it is easy to make fibers with normal or anomalous dispersion, silica fibers predominantly have normal dispersion in the 1.0-1.1 $\mu$m range amplified by Yb, so that dispersion compensation is usually done with a free-space dispersive delay line \cite{Nugent-Glandorf_OptLett2011,Zhou_OptExp2008, Buckley_OptLett2006} or fiber Bragg gratings.\cite{Schibli_NatPhot2008,Hartl_CLEO2005} Fiber Bragg gratings can allow for all-fiber designs, but require very careful design before assembly, as the dispersion is not adjustable. Oscillators with a free-space dispersive delay line allow tuning to find zero dispersion. As we show in section \ref{sec:stabilization}, with transmission gratings, fine-tuning of the grating separation in such a delay line using piezo-electric actuators can also be used to control the comb's carrier-envelope offset frequency.

For mode locking, NPE in fiber\cite{Agrawal_NonlinearFiberOpticsBook,Agrawal_AppNonlinearFiberOptics,Hofer_IEEEJQE1992} provides a fast artificial saturable absorber\cite{Weiner_book2009} that does not require any special components, but is sensitive to temperature or humidity changes. On the other hand, lasers based on real saturable absorbers, such as semiconductor saturable absorber mirrors (SESAM), can be made very environmentally stable, but typically have larger phase and amplitude noise.\cite{Newbury_talk, Sinclair_RSI2015}\footnote{The quiet Yb-oscillators described in references\citenum{Schibli_NatPhot2008} and \citenum{Ruehl_OptLett2010} actually use a combination of both NPE and the SESAM} In our lasers, we use NPE mode locking, and have observed free-running comb-tooth linewidths less than 30 kHz and residual intensity noise less than -130 dBc/Hz for frequencies above 10 kHz (see figure \ref{fig:RIN}). We have also observed reasonable long-term stability in a laboratory setting (more details below).

\begin{figure}[t]
    \centering
   \includegraphics[width=\linewidth]{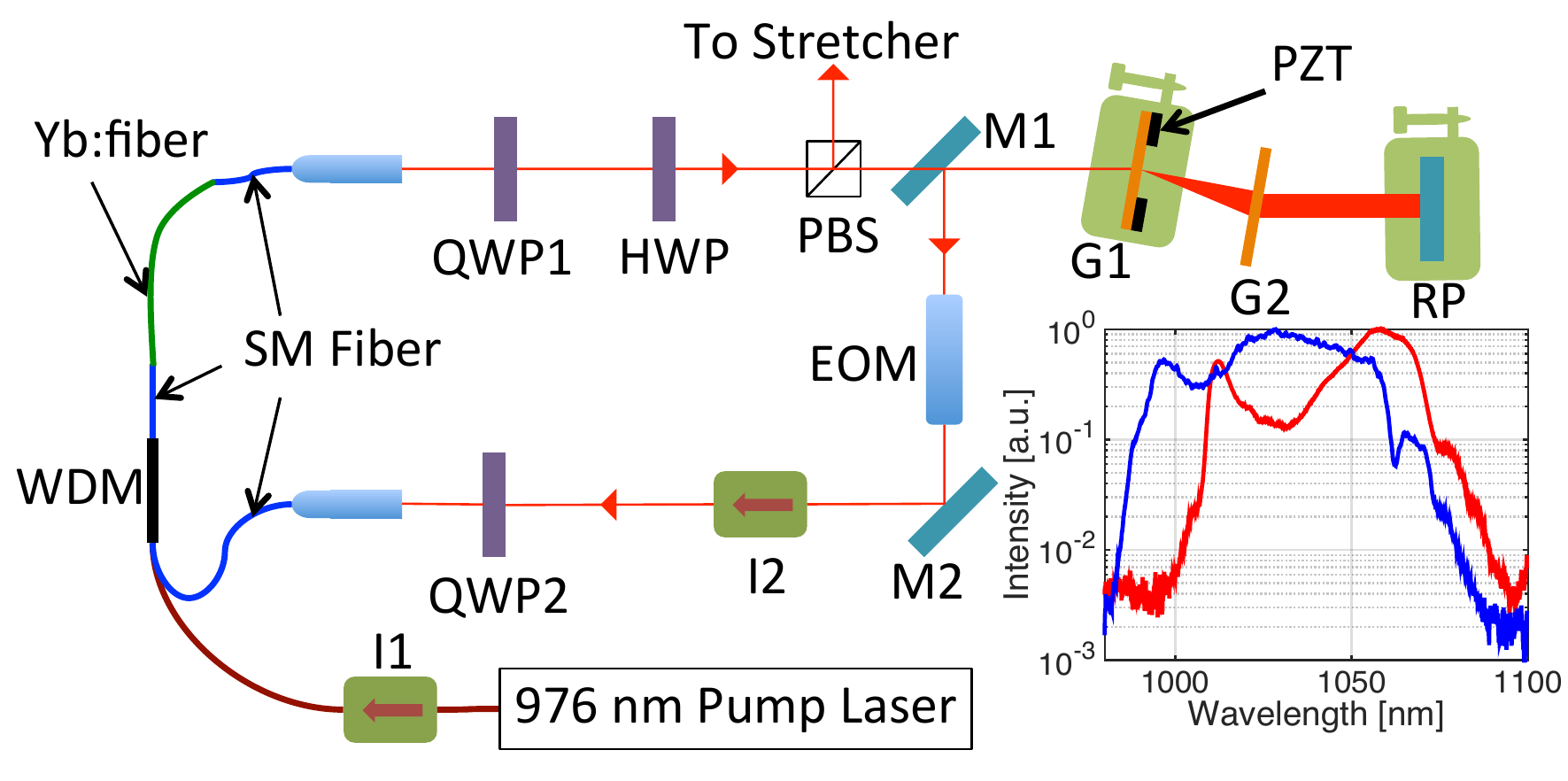}
    \caption{\small \textbf{NPE ring oscillator design}. I = Faraday isolator, WDM = wavelength division multiplexer, QWP = quarter-wave plate, HWP = half-wave plate, PBS = polarizing beam splitter, SM = single-mode, EOM = electro-optic modulator, G = grating, RP = roof reflecting prism, M = mirror. Inset: Typical mode-locked spectra for oscillator A (red) centered at 1060 nm and oscillator B (blue) centered at 1035 nm, both operating near zero net GDD.}
    \label{fig:osc}
\end{figure}

\subsection{Oscillator Construction}

The basic layout for both  Yb:fiber oscillators is shown in figure \ref{fig:osc}. The fiber section provides gain and nonlinearity while the components in the free space section compensate the dispersion of the fiber, manipulate the polarization, and actuate on the pulse's round-trip group delay and carrier-envelope offset phase. The main differences between the two lasers are that (1) the design wavelengths of the components are different to accommodate the different center wavelengths. Oscillator A is designed to operate at 1060 nm and oscillator B is designed to operate at 1035 nm. (2) Oscillator B operates with much lower residual third-order dispersion (TOD) due to a shorter electro-optic modulator (EOM) and larger pitch diffraction gratings (B = 600 groove/mm vs. A = 1000 groove/mm). Complete lists of all the components can be found in the supplemental material, but here we briefly highlight some important aspects.

The pump laser is a fiber Bragg grating stabilized diode laser operating at 976 nm (Oclaro LC96L76P-20R). Basic current and temperature controllers from Thorlabs (LDC210C and TED200C) are used to drive the pump laser. The noise specifications of this diode current controller are sufficient to obtain low-noise operation because the effect of high-frequency pump power fluctuations is suppressed by the low-frequency roll-off of the pump-modulation transfer function observed for these Yb:fiber oscillators, shown in figure \ref{fig:transferfunctions}a). Although the pump laser's fiber is polarization maintaining (PM), we simply splice this onto non-PM fiber for injection into the oscillator cavity with a fused wavelength division multiplexer (WDM). In addition to the WDM, a polarization insensitive isolator (I1) is used to isolate the pump laser from the oscillator light.

The fiber assembly is terminated on each end with anti-reflection (AR) coated angled FC/APC connectors which are then plugged into a fiber coupler lens assembly (Thorlabs PAF-X-5-C). These AR coated fiber tips are obtained simply by splicing the ends of AR coated patch cables (Thorlabs P4-980AR-2) to the ends of the gain fiber and WDM fiber pigtail. The use of connectorized fiber tips enables replacing the fiber assembly with minimal realignment and also allows for the rough alignment of the cavity using another fiber coupled laser, if desired. Single-mode Yb-doped gain fiber can be purchased from a variety of companies (Nufern, Thorlabs, Cor-active, ...) with a variety of Yb doping concentrations. We have used either YB1200-4/125 or YB1200-6/125DC from Thorlabs with similar results. The fiber assembly is spliced together with a basic optical fiber fusion splicer. In our lab, we use a refurbished Ericsson FSU 995FA.

The specifics of the fiber lengths are important for a few reasons, and can be found in the supplemental material. The first is that mode locking depends upon the amount of nonlinearity in the fiber\cite{Ilday_OE2005} and the more fiber there is, particularly following the gain fiber, the easier it is to mode lock. Second, if the AR coated fiber tips become damaged, this is usually due to the gain fiber being too long, and is not remedied simply by reducing the pump power. However, the overall length of the assembly and the relative lengths of the different sections does not have be controlled with high precision. By changing the fiber assembly, we have operated oscillators with repetition rates from 70 MHz to 97 MHz with the same free-space section and obtained similar performance. 

In the free-space section, zeroth-order waveplates are used for polarization control and tuning of the laser, a Faraday isolator ensures uni-directional operation, an EOM enables fast actuation on the effective cavity length, and a pair of transmission gratings is used for dispersion compensation. A polarization beam splitter cube (PBS) is used as an output coupler, reflecting  vertically polarized light out of the cavity.
The first diffraction grating (G1) is mounted on a manual translation stage for finding zero dispersion, and a piezo-electric transducer (PZT) for fine-tuning the comb's carrier-envelope offset frequency. A right angle prism with an AR coated hypotenuse (Thorlabs PS908H-C) serves as the retroreflector in the dispersive delay line, changing the beam height by 5 mm and allowing the beam which initially crossed above to be reflected by D-shaped mirror (M1). In section \ref{sec:stabilization}, we describe the PZT and EOM actuators more carefully using fixed-point analysis.\cite{Newbury_talk, Newbury_JOSAB2007}

For suppression of acoustic noise and mechanical vibration, both oscillators  are enclosed in aluminum sheet metal boxes sided with ``egg-crate" style sound damping foam and built on honeycomb optical breadboards that are supported on the optical table with a 5/8" thick piece of sorbothane rubber. Light is coupled from the oscillator to the amplifier chain via the single-mode fiber pigtail of the stretcher fiber module.

\subsection{Alignment, mode locking, and long-term stability}

Since the stress-induced birefringence of the coiled fiber assembly is unknown, finding the correct positions of the waveplates for NPE mode locking is a somewhat random process. If one simply randomly rotates the three waveplates, this amounts to searching  a three-dimensional space. We have instead developed a reliable procedure for finding mode locking that simplifies the search considerably. First, the oscillator is aligned with the goal of minimizing the pump power necessary for lasing, minimizing the CW lasing threshold, which includes optimizing beam alignment and iterative rotation of polarizers. The pump power is then increased to around 200 mW, well above the minimum pump power that can maintain mode locking, which is about 100 mW. The quarter-wave plate just before the input fiber coupler (QWP2) is rapidly rotated a few degrees back and forth. If mode locking is not achieved, the half-wave plate (HWP) should be stepped a few degrees and then the QWP2 rotation repeated. This process should be repeated until mode locking is achieved, usually within a few iterations. Once the first quarter-wave plate (QWP1) is set for lowest CW lasing threshold, it is generally not necessary to rotate it to find mode locking.

A major problem with NPE Yb:fiber oscillators is that they are prone to multi-pulsing,\cite{Bock_thesis} or the production of more than one pulse circulating in the cavity. The separation between pulses can occur on many different time scales, and thus one needs a range of instruments to detect it. Unlike a Ti:sapphire oscillator, we have frequently observed multi-pulsing to occur with the particularly troublesome separation of 1-500 ps: too short to measure with a typical oscilloscope, but too long to observe easily as interference fringes in the optical spectrum. To be able to detect multi-pulsing at all separations, we employ a combination of three instruments: (1) a low resolution USB optical spectrometer for small pulse separations $<$3 ps, (2) a simple scanning autocorrelator using a GaAsP two-photon photodiode \cite{Ranka_OptLett1997} for the 1-50 ps range, and (3) a fast photodiode (Electro Optics Technology Inc. ET-3010) and sampling oscilloscope (Tektronix 11801C with an SD-26 sampling head, 20 GHz equivalent bandwidth) for longer timescales. A collinear, interferometric autocorrelator is preferred so that one can align it well enough to have confidence in the alignment for longer stage travels, and a two-photon photodiode simplifies the nonlinear signal detection. When multi-pulsing occurs, the first step is to make sure there is nothing terribly wrong with the oscillator. Specifically, check the CW lasing threshold and make sure it is low (typically less than 50 mW). Turn the waveplates, starting with either the HWP or QWP2, until it stops multi-pulsing. 

Once stable mode locking is found, one can search for the lowest noise. The grating spacing for net zero GDD can in principle be calculated using the material parameters for the fibers and other optical elements in the cavity, and this is a good place to start, but it is generally necessary to fine tune this spacing once mode locked. The GDD can be measured using the technique of Knox,\cite{Knox_OL1992} changing the center wavelength either by inserting a knife edge into the dispersed beam or rotating QWP2 slightly (or both). While tuning the grating separation, we monitor oscillator performance using two metrics that can be evaluated quickly: (1) The oscillator relative intensity noise (RIN), measured on a low noise, high bandwidth photodiode, and (2) the free-running heterodyne beat between the oscillator and a narrow linewidth ($<$1 kHz) CW Nd:YAG laser (Innolight Mephisto). Cing\"oz \emph{et al.}\cite{Cingoz_OptLett2011} showed that the phase noise and the RIN are correlated, and as reported by Nugent-Glandorf \emph{et al.}\cite{Nugent-Glandorf_OptLett2011} the laser comb-tooth linewidth depends strongly on the net cavity GDD. Indeed, we have observed optical linewidths ranging between 2 MHz and less than 30 kHz this way, depending on the grating separation.

An important question is: with what precision do I have to find zero GDD? Previously, some authors have emphasized the importance of being slightly normal.\cite{Schibli_NatPhot2008,Fermann_IEEE2009} In our lab, we have observed very similar performance on either side of zero dispersion, within approximately $\pm 2000$ fs$^2$ in both oscillators, in agreement with references \citenum{Nugent-Glandorf_OptLett2011} and \citenum{Song_OE2011}. We have also found that being near zero dispersion is a necessary, but not sufficient, condition to obtaining low-noise performance, and the noise can also depend on the details of NPE and the waveplate angles.

Once satisfactory mode-locked performance is found, we leave the oscillator on indefinitely, and have enjoyed stable hands-free operation for many months at a time in a laboratory setting with reasonable  temperature and humidity control ($\pm$1 $^{\circ}$C, 20-60\% relative humidity). The parts of the laser that in principle have finite lifetimes, the pump diode and the gain fiber, are inexpensive. We have not observed significant degradation of the pump diode performance over three years of nearly continuous operation. However, we have observed that the gain fibers can fail after about one year of continuous operation. The main symptom of this is that the laser just won't mode lock. Lasing thresholds and output powers are similar, but stable mode-locked operation is not re-attained until the gain fiber is replaced, or a new fiber assembly with a fresh gain fiber is installed in the oscillator.

\subsection{Comb Stabilization}
\label{sec:stabilization}

The key element defining an optical frequency comb is that its comb teeth are evenly spaced to an extraordinary precision. \cite{Udem_ICTP2016, Udem_OptLett1999} This occurs naturally in mode-locked lasers and can also occur in other comb-generating systems such as microresonantors \cite{Saha_OptExp2013} and broad-band electro-opticially modulated light fields.\cite{Kourogi_IEEE1993} Once even spacing is established, the comb has two degrees of freedom that determine the  frequencies of the comb teeth. Usually this is expressed in terms of the electronically countable repetition rate $f_{\text{rep}}$ and carrier-envelope offset frequency $f_0$ via the familiar comb formula
\begin{equation}
	\nu_n = n f_{\textss{rep}} + f_{0}
\end{equation}
where $n$ is an integer and the $\nu_n$ are the optical frequencies. Indeed, for self-referenced combs this may be the most sensible parameterization, as it is $f_{\textss{rep}}$ and $f_0$ that are actively controlled. However, for optically referenced combs, combs coupled to cavities, or when discussing the effects of actuators or noise sources, the discussion is often simplified by using a fixed point analysis,\cite{Kuse_OE2015,Newbury_JOSAB2007,Newbury_talk}  writing the comb's optical frequencies as
\begin{equation}\label{eqn:fixedpointdef}
	\nu_n = (n-n^{*})f_{\textss{rep}} +\nu_{n^{*}}
\end{equation} 
where $n^*$ is an integer representing a fixed point of the frequency comb that does not change due to a particular perturbation such as noise or intentional actuation on the laser. In the fixed point picture, one considers the comb teeth simply expanding and contracting around the fixed point via changes in $f_{\textss{rep}}$. The larger the frequency difference is between the fixed point and a particular comb tooth, the more the frequency of that comb tooth changes due to the perturbation.

Since the frequency comb has two degrees of freedom, one needs two feedback loops and two actuators to stabilize the comb. Ideally, these two feedback loops would have zero cross-talk. For example, if one directly stabilizes $f_{\textss{rep}}$ and $f_0$, ideally one actuator would actuate only on $f_{\textss{rep}}$ and the other only on $f_0$. In practice, this goal is almost never reached exactly, which is acceptable as long as one feedback loop can be significantly slower than the other, such that the faster loop can adiabatically track and correct for the cross-talk from the competing loop.

For coupling a frequency comb to an optical cavity, or locking the frequency comb to another optical reference, it is desirable to have one fast actuator with its fixed point far from the optical frequency and another actuator with fixed point near the optical frequency. This allows the fast actuator to have large travel at optical frequencies that are being stabilized, and the second actuator to simply cause the comb to breath around this locked point. 

\begin{figure}[t]
    \centering
   \includegraphics[width=8cm]{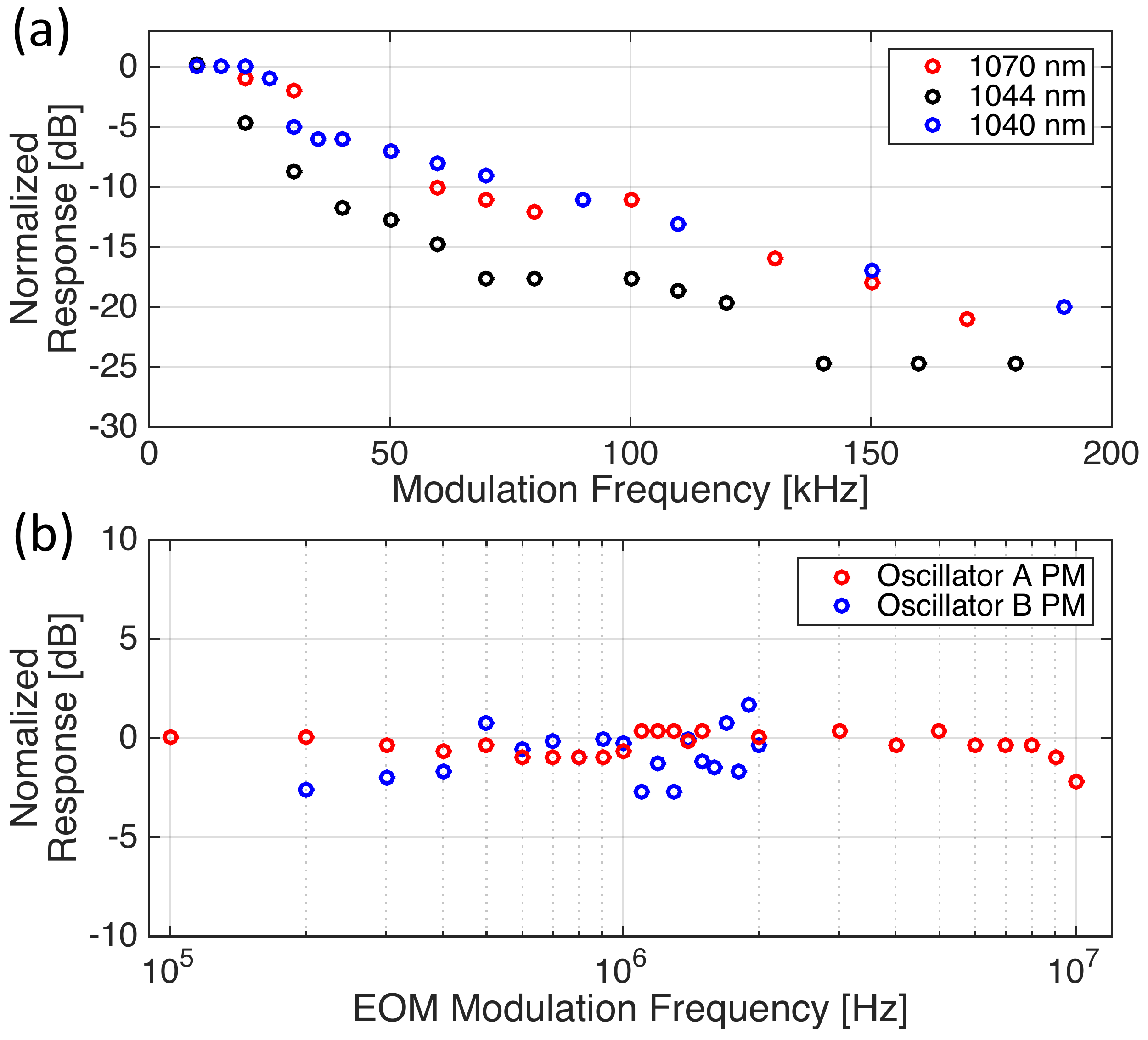}
    \caption{\small \textbf{Transducer transfer functions}. a) Pump amplitude to laser amplitude modulation transfer function. b) Voltage to phase modulation transfer functions for the intracavity EOMs in oscillators A and B.}
    \label{fig:transferfunctions}
\end{figure}

A commonly used combination of actuators to accomplish this is a fast intracavity EOM \cite{Hudson_OL2005,Zhang_IEEE2012,Iwakuni_OE2012,Swann_OE2011,Nakajima_OE2010,Benko_OptLett2012,Lee_OL2012} for cavity length changes, with fixed point near DC ($n^{*}\sim 0$), and the pump laser power for changing the intracavity pulse's round trip phase shift, with fixed point near the optical carrier frequency.\cite{Newbury_talk, Newbury_JOSAB2007, Hellwig_OL2014} While fast bandwidth can be obtained via actuating on the pump current in some laser designs,\cite{Cingoz_OptLett2011} the bandwidth attainable with this actuator depends on the details of the laser and population inversion dynamics.\cite{Bao_Ol2014} In the current ring cavity design with lower loss and smaller gain than the laser in reference \citenum{Cingoz_OptLett2011}, and thus lower relaxation oscillation frequency, we have observed the bandwidth of pump power modulation to be quite limited, as shown in figure \ref{fig:transferfunctions}a). To record this data, we modulate the pump diode current sinusoidally and record the amplitude modulation on the comb light with a photodiode and a spectrum analyzer. The pump current to pump power transfer function (not shown) was independently verified to be flat out past 1 MHz modulation frequencies with this setup, confirming that the roll-off is due to transfer function of the Yb laser. 

In our lasers we use bulk EOMs for fast (multi-MHz bandwidth) actuation with a fixed point near DC, and instead of the pump power we  use $\mu$m-scale piezo-electric adjustments of the grating spacing, which we show below has a fixed point near the optical frequency. 

For the EOM, we use simple bulk-crystal EOMs. Previous authors \cite{Benko_OptLett2012} have used short (few mm) EOM crystals due to concern over higher order dispersion. However, short crystals require multi-kV voltages to produce substantial phase shifts, and this is challenging to provide with high bandwidth. We have found that the remarkable tolerance of Yb:fiber oscillators to uncompensated higher-order dispersion\cite{Kuse_OL2010} enables the use of longer crystals with correspondingly lower voltage requirements. Oscillator A uses a commercial phase modulator (Thorlabs EO-PM-NR-C2) with a 40 mm MgO doped Lithium Niobate crystal and a $V_{\pi}$ of 250 V, and oscillator B uses a home-built EOM with a 4 mm LiTaO$_3$ crystal (United Crystals) compressed between brass and acrylic fixtures\cite{Hyodo_OptComm1999} and a $V_{\pi}$ of approximately 3 kV. As shown in figure \ref{fig:transferfunctions}b), both EOMs allow phase modulation with multi-MHz bandwidth without piezo-elastic resonances that have limited previous efforts.\cite{Benko_OptLett2012} To record this data, we drive the EOM with a sinusoidal  voltage and record the amplitude of phase-modulation side bands on the heterodyne beat with the CW Nd:YAG laser, taking care of the fact that the intracavity EOM modulates the laser's \emph{frequency} but the sidebands on the beat report on the \emph{phase} modulation depth.

In both lasers we  supply a 2 MHz sinusoidal  voltage to the EOM to put frequency modulation (FM) sidebands on the comb that enable Pound-Drever-Hall (PDH) locking of the combs to passive optical cavities.\cite{Jones_OptLett2004, Black_AmJPhys2001,Drever_APB1983} We measured the residual amplitude modulation (RAM) on the output light of the oscillator to be less than -90 dBc when driving the EOMs with a 20 V (peak to peak) sine wave (more than what is typically required for PDH locking). The EOM alignment can be fine-tuned \emph{in situ} by minimizing this RAM.

For actuation on the grating spacing, the first grating is glued to a ring PZT (Noliac NAC2125) using Loctite Hysol 1C-LV epoxy (also sold under the trade name Torr-seal), and this allows $ > 10$ kHz of bandwidth before encountering mechanical resonances. Shifting the carrier-envelope offset phase of a pulse by actuation on the grating separation of a pulse compressor has been employed for carrier-envelope offset phase stabilization in amplified Ti:sapphire lasers (after the amplifier chain),\cite{Li_OptLett2006} but to our knowledge this is first report of doing this inside a laser cavity. Here we derive the resulting frequency shifts for a transmission grating geometry and show that for transmission gratings operated in Littrow condition, the fixed point is at the optical carrier frequency $\nu_{\textss{optical}}$, such that the change in $f_0$ is approximately $\nu_{\textss{optical}}/f_{\textss{rep}}$ larger than the change in $f_{\textss{rep}}$. 

For the parallel grating pulse compressor illustrated in figure \ref{fig:gratings}, the total phase shift for one pass through the grating pair is given by:\cite{Treacy_IEEE1969}
\begin{equation}\label{eqn:gratingphase}
	\phi_{g}(\omega) = \frac{\omega}{c}p(\omega) - \frac{2\pi}{d}G\tan(\beta)
\end{equation}
where $\omega=2\pi\nu$ is the angular frequency, $\phi_{g}(\omega)$ is the spectral phase, $p(\omega)$ is the frequency dependent optical path length through the compressor, $\beta$ is the angle of diffraction determined from the grating equation, $\sin(\alpha) + \sin(\beta) = \lambda/d$, with $\alpha$ the angle of incidence measured from normal, $d$ is the grating pitch, $\lambda$ is the wavelength, and $G$ is the distance between the gratings measured perpendicular to the grating surfaces. The second term in equation (\ref{eqn:gratingphase}) accounts for the $2 \pi$ phase shift encountered by the light for each grating groove traversed and must be included to obtain correct results.\cite{Treacy_IEEE1969} Careful inspection of the angle-dependent path length shows that
\begin{equation}\label{eqn:dphidG}
	\frac{d \phi_{g}}{dG} = \frac{\omega}{c} \left( \frac{1}{\cos(\beta)} - \frac{\cos(\alpha+\beta)}{\cos(\beta)} \right) - \frac{2\pi}{d}\tan(\beta)
\end{equation}
Now one is tempted to locate the fixed point, $\omega^*$, by setting equation (\ref{eqn:dphidG}) equal to zero and solving for $\omega$, but this is not generally correct because mode locking demands that the comb teeth remain evenly spaced, and thus the differential phase shift between comb tooth $n+1$ and $n$ must be the same as the differential phase shift between comb tooth $n+2$ and comb tooth $n+1$. Enforcing this fact that the comb has only two degrees of freedom amounts to linearizing the spectral phase using the phase shifts obtained near the optical carrier frequency, viz.
\begin{equation}\label{eqn:dphicombdG}
	\frac{d\phi_{\textss{comb}}}{dG} =  \left. \frac{d \phi_{g}}{dG} \right|_{\omega_0} + \left. \frac{d \tau}{dG} \right|_{\omega_0}(\omega - \omega_0)
\end{equation}
 where $\tau = d\phi_{g}/d\omega$ is the frequency dependent group delay, which is evaluated at the optical carrier frequency $\omega_0$ in equation (\ref{eqn:dphicombdG}). The fixed point is then given by
\begin{equation}\label{eqn:omegafixed}
	\omega^* = \omega_0 -  \left( \left. \frac{d \phi_{g}}{dG} \right|_{\omega_0} \right) \left({\left. \frac{d \tau}{dG} \right|_{\omega_0}} \right)^{-1}
\end{equation}
At the Littrow condition, $\alpha = \beta(\omega_0) = \sin^{-1}(\pi c/ \omega_0 d)$, one can show that the phase shift due to changing the grating separation, $d \phi_{g}/dG$, is identically zero and the fixed point is thus at the optical carrier frequency. For the more realistic scenario that the gratings end up slightly off-Littrow, one can use equations (\ref{eqn:fixedpointdef}), (\ref{eqn:dphidG}), and (\ref{eqn:omegafixed}) along with the relation $d f_{\textss{rep}}/dG = -f_{\textss{rep}}^2 \left. d\tau/dG \right|_{\omega_0}$ in order to determine the changes in comb tooth frequencies. One can also derive relations for the changes in $f_{\textss{rep}}$ and $f_0$. For two passes through the grating pair, under Littrow conditions, the result is: 

\begin{figure}[t]
    \centering
   \includegraphics[width=8cm]{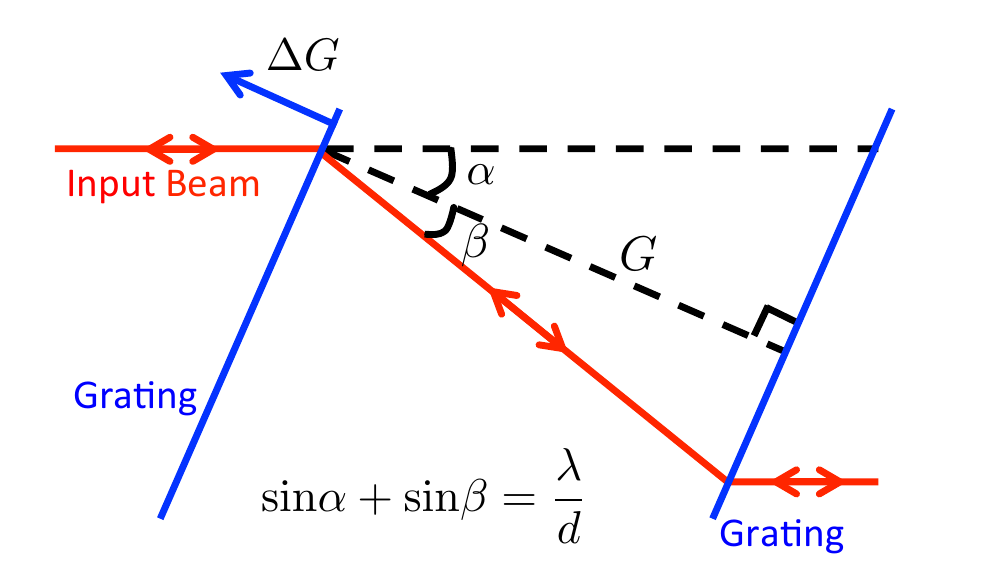}
    \caption{\small \textbf{Transmission grating geometry}. Illustration of the notation for the intracavity grating compressor. The grating spacing is adjusted slightly ($\Delta G$) using a PZT to control the comb's carrier-envelope offset frequency. The arrow indicates the direction of positive $\Delta G$.}
    \label{fig:gratings}
\end{figure}



\begin{equation}
\label{equ:frepdg}
\frac{df_{\textss{rep}}}{dG}=~-\frac{4\pi f^2_{\textss{rep}}}{ \omega_o d}\tan(\alpha).
\end{equation}

\begin{equation}
\label{equ:fodg}
\frac{d f_0}{dG}~= -n^* \frac{df_{\textss{rep}}}{dG} = \frac{2 f_{\textss{rep}}}{d}\tan(\alpha).
\end{equation}

\begin{figure}[t]
    \centering
   \includegraphics[width=8cm]{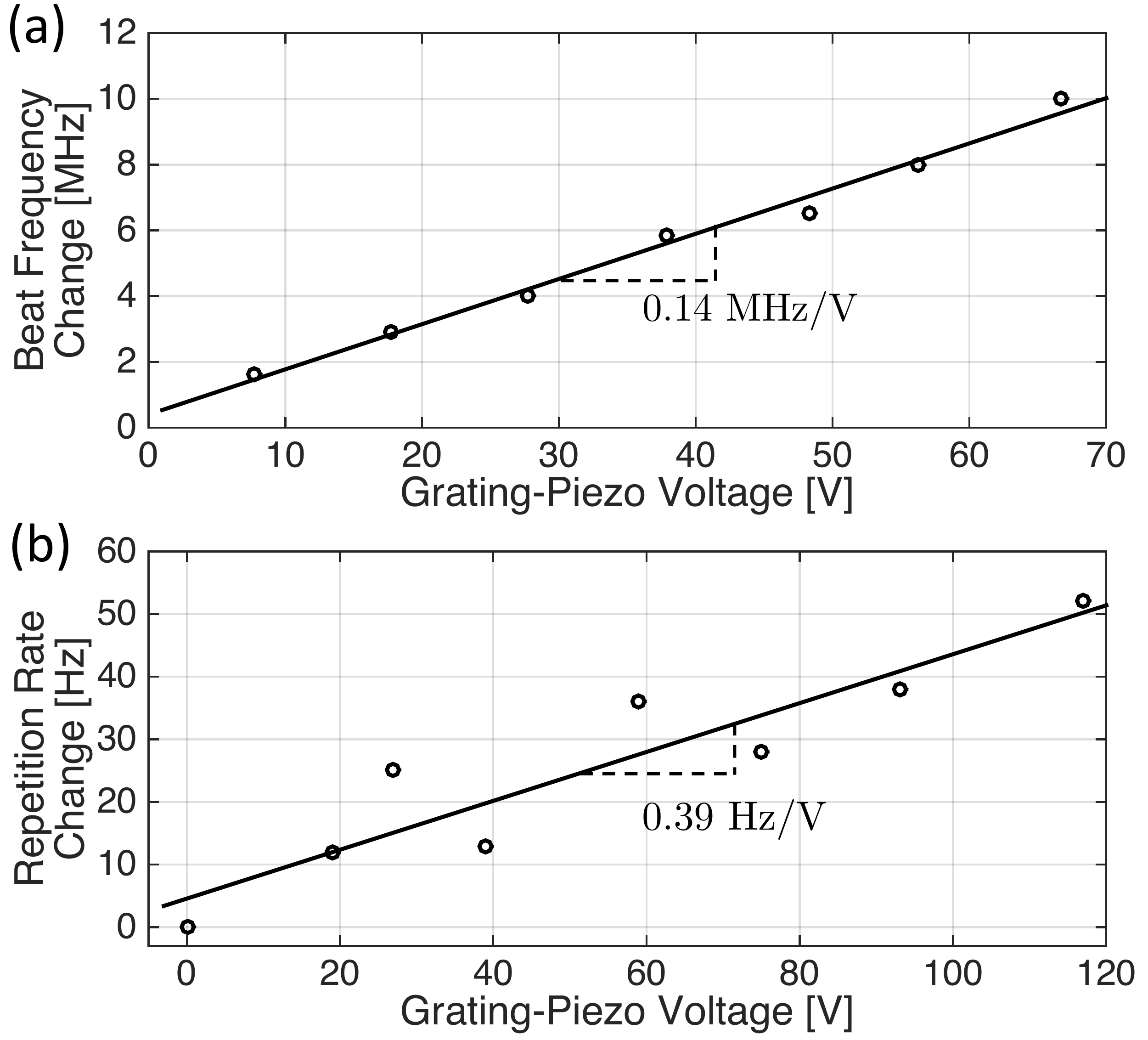}
    \caption{\small \textbf{Grating acutation}. a) 1064 nm beat frequency change with grating separation. b) Repetition rate change with grating separation. The data indicate that the fixed point is near the optical frequency.}
    \label{fig:grating}
\end{figure}

We have experimentally verified this analysis by recording changes in the comb repetition frequency and an optical comb tooth near 1064 nm with oscillator A when voltage is applied to the grating PZT. The data are shown in figure \ref{fig:grating}. The repetition rate changes are measured using a photodiode and a frequency counter. The changes in the optical frequency are measured by recording the beat frequency of an unstabilized heterodyne beat between the comb and the CW Nd:YAG laser. Linear fits to the data give slopes of $\frac{d \nu_{\textss{beat}}}{dV} = 0.14 $ MHz/V and $\frac{d f_{\textss{rep}}}{d V} = 0.39$ Hz/V. The number of comb teeth between 1064 nm and the fixed point can then be simply calculated from the ratio $\Delta \nu_{\textss{beat}} / \Delta f_{\textss{rep}} = 3.3 \times 10^5$. So the fixed point lies only approximately 30 THz away from the optical carrier frequency of 283 THz. This is consistent with the above equations and $\alpha$ deviating from the Littrow angle by approximately 4 degrees, which is realistic given our ability to initially set the grating angle in the laser and the $\alpha$ dependence of the grating's diffraction efficiency.

The grating can also be used to make a shift purely in $f_0$, if the grating is moved parallel to its surface, similar to the motion of a sound wave in an acousto-optic frequency shifter. Here the phase shift is simply $2 \pi$  per grating pitch moved,\cite{Wise_PRL2005} and the fixed point is at $\nu = \infty$. Using both parallel and perpendicular motions, in principle one could completely control the comb with only $\mu$m-scale motions of the grating alone, as the two motions have different fixed points.

\section{Chirped-Pulse Amplification in Large Mode Area Photonic Crystal Fibers}\label{sec:Amplifiers}

Amplification of continuous wave lasers to high average power in Yb:fiber is straightforward, but amplification of femtosecond pulses presents additional complications. The long length of fiber presents a large amount of dispersion  even for large mode area (LMA) fibers, and it is much more difficult to avoid accumulated nonlinear phase shifts  than in bulk solid-state lasers. Designers  of ultrafast fiber lasers usually take one of two approaches: embrace nonlinearity \cite{Fermann_PRL2000, Dudley_NatPhys2007, Zhao_OptExp2014, Zhou_OptExp2005, Liu_OptLett2015} or use stretchers and compressors with very large GDD to avoid it.\cite{Ruehl_OptLett2010, Roser_OptLett2005, Zhao_AppPhysExp2016, Eidam_OptLett2010,Wunram_OptLett2015} For comb applications, linear amplification, in which the $B$-integral, or accumulated nonlinear phase shift throughout the amplifier chain, is less than one, is generally preferred because then the amplified comb's coherence properties are determined mainly by the oscillator. In nonlinear amplification, amplitude noise from the high-power pump diodes in the amplifier chain could write phase noise on the amplified comb,\cite{Ruehl_OPN2012,Fermann_IEEE2009} although we are aware of some recent efforts using high-power nonlinear fiber amplification for comb applications.\cite{Mills_SPIE2015,Wu_FiO2013}

In our lasers, we have used linear CPA, but have strived to maintain modest stretcher/compressor dispersion by (1) seeding the amplifiers with very broad spectra and (2) maximizing the mode area of the seed light throughout the amplifier chain, even if this means seeding amplifiers below saturation. For simplicity and low-cost, we make use of fiber stretchers based on anomalous third-order dispersion depressed cladding fibers (OFS) and grating compressors based on inexpensive polymer transmission gratings (Wasatch Photonics). While we do not quite reach transform limited pulses on either laser system, the benefits of the simplicity of this scheme have outweighed the slightly reduced performance. 

\subsection{9 W amplifier at 1060 nm} \label{sec:10Wamp}

\begin{figure}[t]
    \centering
   \includegraphics[width=0.9\linewidth]{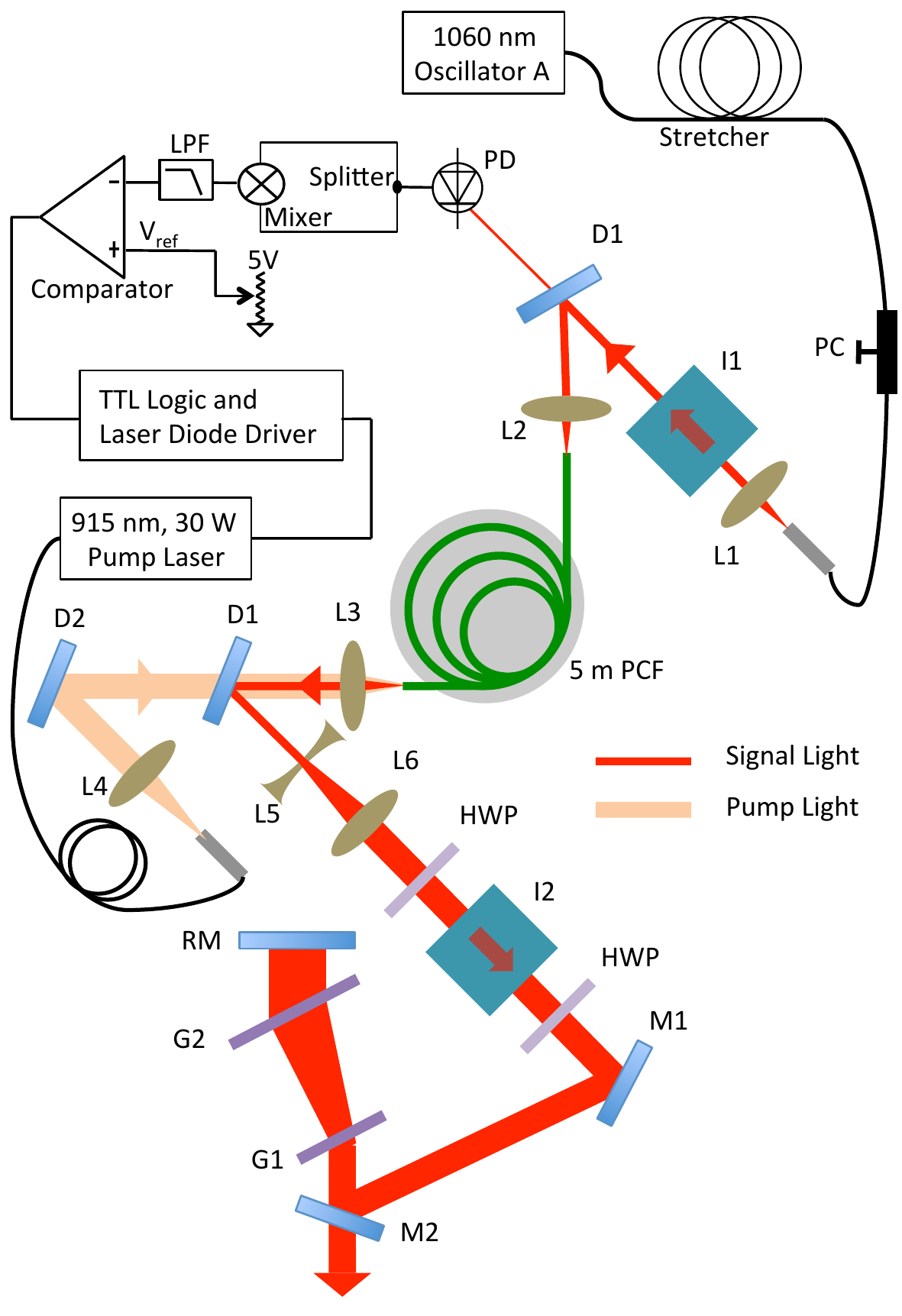}
    \caption{\small \textbf{9 W amplifier layout}. Pulses from oscillator A are stretched and amplified in a 5 m flexible PCF amplifier up to 9 W average power without pre-amplification. Component lists can be found in the supplemental material. PC = polarization controller. D = dichroic mirror. L = lens. PD = photodiode. RM = roof reflecting mirror. LPF = low pass filter. V$_{\mathrm{ref}}$ = reference voltage.}
    \label{fig:10Wamp}
\end{figure}


\begin{figure}[t]
    \centering
   \includegraphics[width=0.89\linewidth]{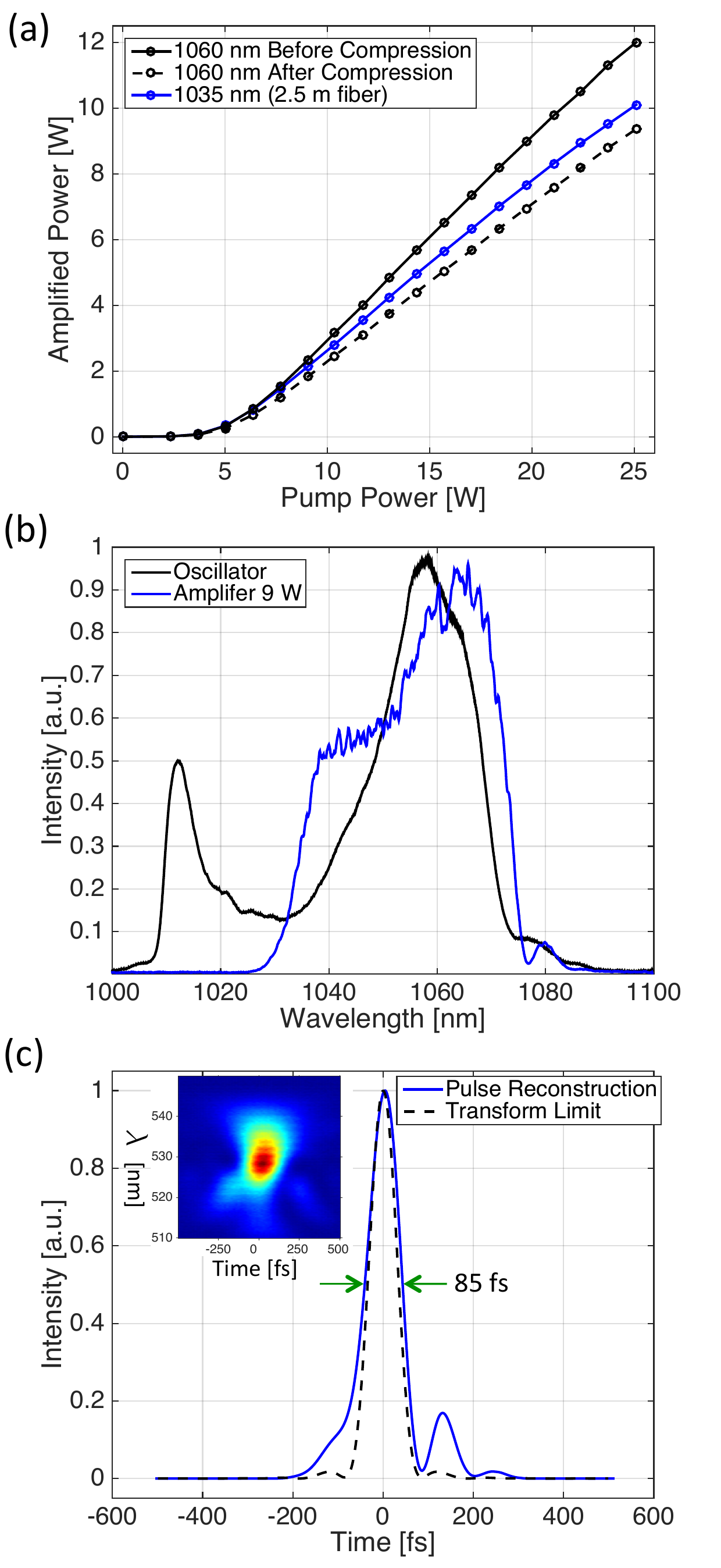}
    \caption{\small \textbf{Flexible PCF amplifier performance}. a) Black: Output power of the 5 m, 1060 nm PCF amplifier, before and after the compressor vs. 915 nm diode pump power. Blue: output power of the 2.5 m, 1035 nm PCF amplifier. b) Laser spectra measured from the oscillator and compressor output. c) Raw FROG trace (inset) and retrieved pulse shape of the compressed pulses compared to the transform limit calculated from the spectrum in b).}
    \label{fig:1060result}
\end{figure}

A schematic of the 9 W CPA scheme is shown in figure \ref{fig:10Wamp}. The light from oscillator A is coupled into a stretcher fiber module custom made by OFS Specialty Photonics Division with FC/APC connectorized SMF-980 fiber pigtails. Between the stretcher and the following Faraday isolator (I1), about half the seed power is lost. After the dichroic mirror (D1), which is used to isolate the pump light and seed (or signal) light, 15-20 mW is launched into the 5 m amplifier fiber. The amplifier fiber is a doped, LMA (760 $\mu$m$^2$) PCF terminated with sealed ends and copper SMA 905 connectors, purchased from NKT Photonics (aerogain Flex 5.0). This fiber is end pumped with a 30 W, 915 nm pump diode (nLight Element). After the second dichroic mirror (D2), the beam is expanded to 4.2 mm ($1/e^2$ diameter), sent through another Faraday isolator, and compressed using a pair of polymer transmission gratings (Wasatch Photonics) and a roof reflector (M3). Below, we discuss these features in more detail and the design decision processes behind them. 

The coiled amplifier fiber is supported on a circular aluminum plate. The pump end, where the optical power is the highest, is mounted in a water-cooled copper clamshell assembly. Detailed drawings of the copper clamshell can be found in the supplemental material (figure S3). The seed end of the amplifier fiber is screwed into an SMA connector (Thorlabs HFB001) mounted on a flexure stage (Thorlabs MicroBlock MBT616D). The heavy copper mode-stripper assembly of the fiber is further supported by shims placed on the flexure stage. The output of the stretcher fiber and pump diode fiber are mounted on flexure stages in similar fashion. For delivering the pump light, the pump diode fiber pigtail is spliced onto the end of an AR coated multimode patch cable (Thorlabs M105L02S-B). The pump light is launched counter-propagating to the amplified seed light to reduce the accumulated nonlinear phase shift ($B$-integral) in the amplifier fiber. Despite not being ``all-fiber", we have observed consistent performance from this mechanical setup without alignment for more than 2 years of operation.

The saturation power for the LMA PCF amplifier is more than 200 mW, and many previously reported amplifier systems using these PCF amplifiers have employed a fiber pre-amplifier with smaller mode area between the oscillator and the PCF amplifier \cite{Ruehl_OptLett2010, Wunram_OptLett2015, Zhao_AppPhysExp2016,Mills_SPIE2015,Wu_FiO2013} in order to seed the power amplifier at saturation. Since we are seeding the amplifier well below the saturation power, a threshold-like behavior is observed in the amplified power vs. pump power curves shown in figure \ref{fig:1060result}a), reducing the efficiency of the amplifier. While this one-stage amplification scheme is less efficient, it is much simpler due to (1) the lack of all the pre-amplifier components and (2) the pulses do not have to be stretched as much to avoid nonlinearity, since all the high power propagation is done in LMA fiber. A smaller stretching/compression ratio allows for looser tolerances on matching the higher-order dispersion of the stretcher and compressor. The 5 Watts of pump power wasted before the amplifier reaches saturation are not really of consequence, due to the low cost of high-power pump diodes.

\begin{table}[b]
\begin{tabular} {c | c | c | c}
Component & GDD (ps$^2$) & TOD (fs$^3$) & notes\\ 
\hline
Stretcher &  1.65 & -7.9$\times$10$^6$ & \\
Amplifier & 0.095 & 2.06$\times$10$^5$ & FS, L$_{\mathrm{tot}}$=5 m \\
Isolators	  & 0.011 & 6.6$\times$10$^3$ & TGG, L$_{\mathrm{tot}}$=8 cm	\\
Compressor &  -1.76 & 7.63$\times$10$^7$ & \\
\end{tabular}
\caption{Dispersion budget for the 1060 nm laser. FS = fused silica. TGG = Terbium Gallium Garnet.}
\label{table:10Wdispersion}
\end{table}

Two concerns with underseeding the amplifier are (1) noise due to amplified spontaneous emission (ASE),\cite{Agrawal_NonlinearFiberOpticsBook} and (2) catastrophic damage to the amplifier fiber due to self-lasing and Q-switching. Regarding (1), despite the expectation of increased ASE, the measured RIN spectrum of the amplified light (see pre-amplifier curve in figure \ref{fig:RIN}) indicates that the main source of noise on the amplified light is due to the pump diode RIN, not ASE. Regarding (2), while we do not know what the lowest necessary seed power is to avoid catastrophic damage, we can say that we have run these PCF amplifiers (both at 1035 nm and 1060 nm) for many hours with seed powers as low as 10 mW without observing damage. We continuously monitor the seed light with a fast (100 MHz bandwidth) photodiode (PD) and a simple interlock circuit, shown in figure \ref{fig:10Wamp}, which immediately shuts off the pump diode in the event that the RF power from the photodiode drops below a set threshold, indicating reduced power or loss of mode locking. Pump diode drivers from VueMetrix Inc. shut off in less than 50 microseconds upon receiving an electronic signal, much shorter than the energy storage time in Yb of approximately 1 ms. \cite{Paschotta_IEEE1997} A similar interlock system is used for the amplifiers of the 80 W laser discussed in section \ref{sec:80Wamp}.


Another feature of this amplifier system to note is the operating wavelengths. The amplified light is at 1060 nm and the pump light is at 915 nm, whereas most ultrafast Yb amplifiers are pumped at 975 nm and amplify light at 1030-1040 nm, where the absorption and emission cross sections are largest.\cite{Paschotta_IEEE1997} We use 915 nm for the pump wavelength because the absorption feature at 975 nm is  narrow, requiring tight control over the pump diode wavelength for efficient pumping. While the absorption cross section at 915 nm is three times lower, the absorption maximum there is also much broader, which loosens the requirements for controlling the pump diode wavelength, and thus temperature, considerably. With the long 5 m PCF, more than 90\% of the pump light is still absorbed. For the amplified wavelength, we operate this 9 W laser at 1060 nm, far to the red of the Yb emission maximum, because one can amplify  with considerably less gain narrowing, and this has also been employed in a few other linear CPA designs. \cite{Ruehl_OptLett2010,Schibli_NatPhot2008} Figure \ref{fig:1060result} shows the output spectrum of the amplified laser with more than 30 nm of bandwidth.

\begin{figure*}[t]
    \centering
   \includegraphics[width=0.9\linewidth]{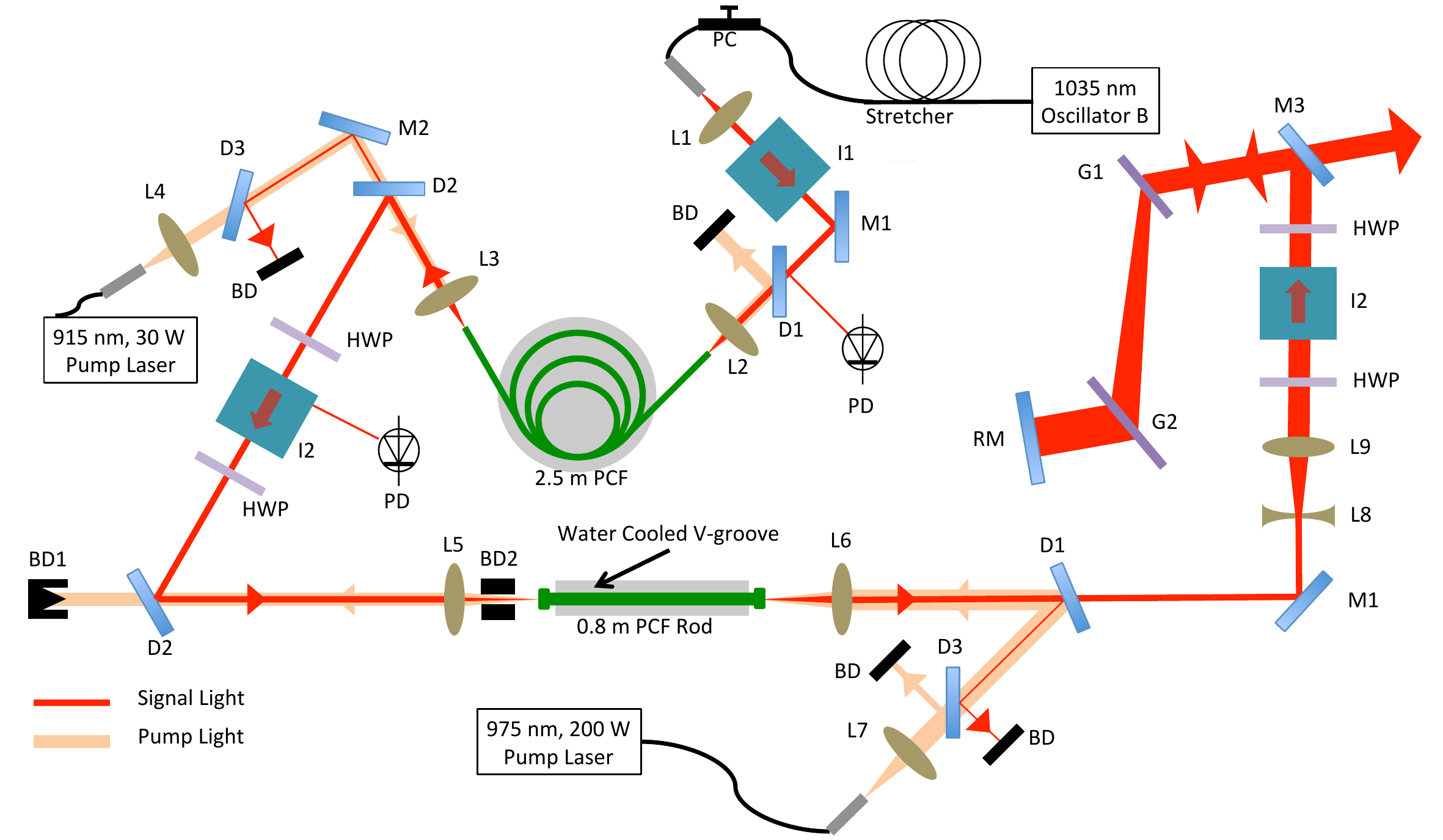}
    \caption{\small \textbf{80 W amplifier layout}. Chirped pulses are amplified first in a 2.5 m flexible photonic crystal fiber and then in a 0.8 m rod fiber. Component lists can be found in the supplemental material.}
    \label{fig:80Wamp}
\end{figure*}

A dispersion budget for the CPA system following oscillator A is shown in table \ref{table:10Wdispersion}. The oscillator pulses are stretched to approximately 100 ps duration in the fiber stretcher module and compressed to $<$100 fs after amplification using a Treacy-style compressor  with 1250 groove/mm transmission gratings (Wasatch Photonics). We measure an overall compressor efficiency of 77\%, corresponding to a diffraction efficiency of $(77\%)^{1/4}$  = 94\%. Figure \ref{fig:1060result}c) shows a second harmonic generation (SHG) frequency resolved optical gating (FROG) trace taken with the laser at full power using a commercial FROG system (Mesa Photonics FROGscan Ultra). While a free-space Offner-type stretcher\cite{Offner_Patent,Cheriaux_OptLett1996} would allow for more tunability than the fiber stretcher module, and perhaps better compensation of higher order dispersion,\cite{Kane_JOSAB1997} it would also add substantial mechanical complexity and cost. The fiber stretcher module is alignment free, and we have observed nearly transform-limited performance in both CPA systems using these for these stretcher modules.

\subsection{80 W Amplifier at 1035 nm}\label{sec:80Wamp}


Figure \ref{fig:80Wamp} shows a higher power laser capable of 80 W average power via two-stage amplification. The PCF pre-amplifier is similar to the system described in section \ref{sec:10Wamp}, except that a shorter 2.5 m fiber is used due to the larger gain and absorption of Yb at 1035 nm. For the second stage, a 0.8 m long Yb-doped PCF rod (NKT Photonics, aeroGAIN-ROD-PM85-POWER) with a 3,400 $\mu$m$^2$ mode field area, end pumped by a 200 W pump diode module (LIMO, LIMO200-F200-DL980-S1886) is used for the rod amplifier. When seeding the rod with 6.4 W, 96 W emerges with excellent beam quality when pumping the rod with 200 W of pump power. The large mode field diameter of the rod allows the laser to maintain linear amplification to $\mu$J pulse energies using only a modest stretcher dispersion of 5.3 ps$^2$, or stretched pulse durations of only approximately 300 ps. 

\begin{figure}[t]
    \centering
   \includegraphics[width=0.89\linewidth]{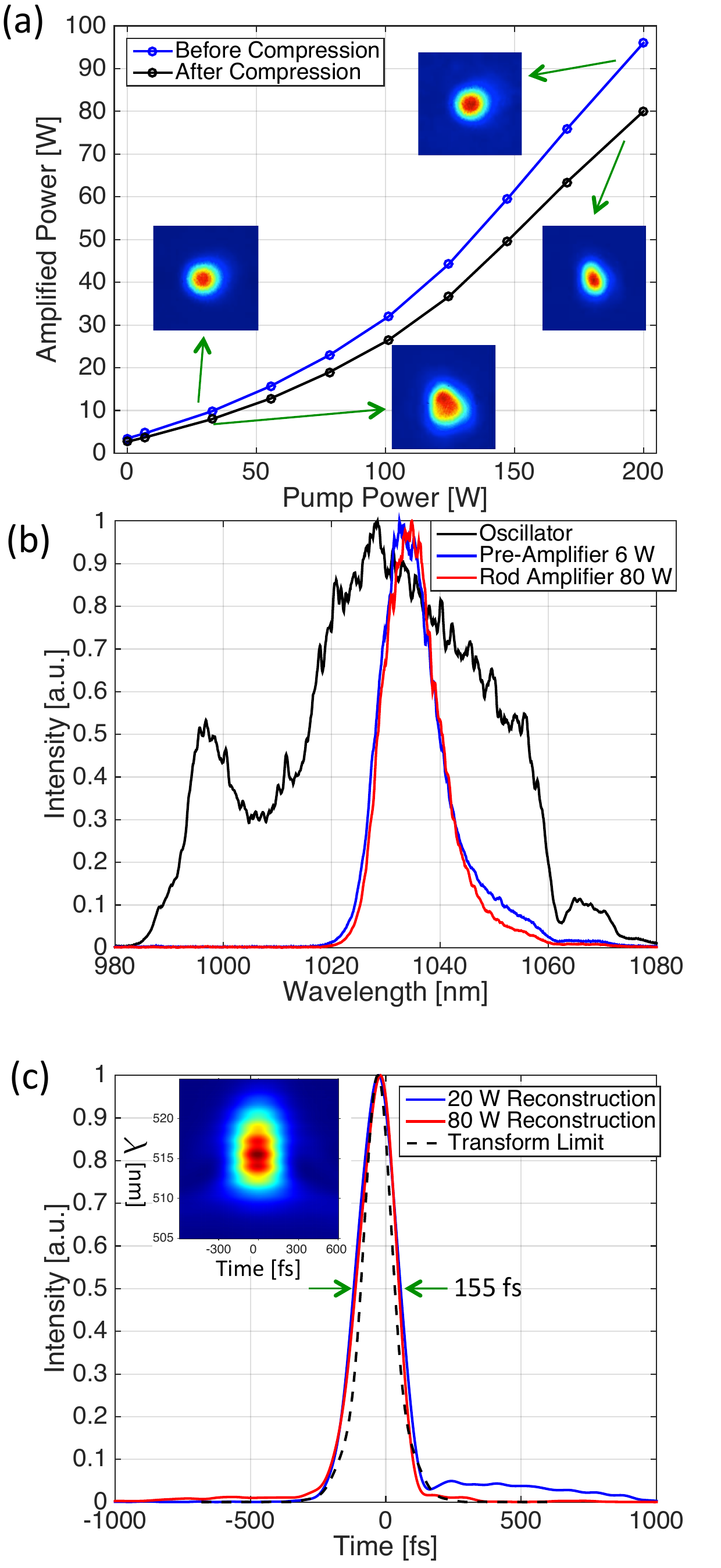}
    \caption{\small \textbf{Rod amplifier performance}. a) Output power of the Yb:fiber rod amplifier, pumped by a counterpropagating 975 nm pump diode. The beam mode at lower power and full power is shown for both the output of the rod fiber and the compressor. b) Laser spectra measured from the oscillator and amplifier output. c) Raw FROG trace at 80 W (inset) and retrieved pulse shapes of the compressed pulses compared to the transform limit calculated from the spectrum in b) for both low and high power operation.}
    \label{fig:80Wresult}
\end{figure}

The available rod fibers are shorter than flexible PCF, necessitating pumping at 975 nm and amplification in the more conventional 1030-1040 nm region to achieve efficient gain. With the amplified center wavelength at 1035 nm, we observe significant gain narrowing\cite{Kuznetsova_OSA2007,Siegman} in the amplifier chain, as shown in figure \ref{fig:80Wresult}b). However, we are still able to compress the rod amplifier output to very clean 155 fs pulses, as shown in figure \ref{fig:80Wresult}c). Pumping at 975 nm requires tighter control over the pump diode wavelength, increasing the cost of the pump diode. In our system we do not actively control the temperature, and this is the reason for the nonlinear amplified  power vs. pump power curves of figure \ref{fig:80Wresult} - the pump laser wavelength is changing as the power is increased (see figure S2 in the supplemental material). 


The rod fiber is supported along its length by a water-cooled aluminum V-groove, held loosely by only two pieces of kapton tape. The end caps, the parts that are critical for alignment, are rigidly located in V-shaped jaws mounted on the optical table separate from the V-groove. The high pump power of the rod amplifier can bring complications. When pumped with 200 W, more than 40 W of pump light comes out from the seed end of rod fiber with NA=0.5, which could heat the mount of the lens L5 or the mirror mount for D2, causing a dangerous drift in the seed light alignment. We dump the pump light safely using two water-cooled black-anodized aluminum beam dumps, the annular BD2 to protect the lens mount and BD1 to collect the pump light transmitted through the dichroic mirror (D2). Mechanical drawings for BD2 can be found in the supplemental material.

\begin{table}[b]
\begin{tabular} {c | c | c | c}
Component & GDD (ps$^2$) & TOD (fs$^3$) & notes\\ 
\hline
Stretcher &  5.27 & -2.16$\times$10$^7$ & \\
Amplifier & 0.063 & 1.36$\times$10$^5$ & FS, L$_{\mathrm{tot}}$=3.3 m \\
Isolators	  & 0.017 & 9.9$\times$10$^3$ & TGG, L$_{\mathrm{tot}}$=12 cm	\\
Compressor & -5.32 & 2.15$\times$10$^7$ & \\  
\end{tabular}
\caption{Dispersion budget for the 1035 nm laser. FS stands for fused silica. TGG = Terbium Gallium Garnet.}
\label{table:80Wtable}
\end{table}

Another problem with high average power is thermal lensing and distortions of the beam quality. As shown in figure \ref{fig:80Wresult}, an excellent, nearly Gaussian, spatial profile is observed in the amplified light from the rod amplifier at both low and high power. However, even at low power, we observe distortions of the beam after four diffractions from the inexpensive polymer gratings. At higher power, despite the large expanded beam size of 5.8 mm $1/e^2$ diameter after the telescope (L8 and L9), thermal lensing is observed, and the output mode measured 1.8 m after the compressor, is significantly smaller. Inspection of the beam at various points indicates that the thermal lensing occurs either in this telescope or the Terbium Gallium Garnet crystal of the optical isolator. However, despite the obvious thermal lensing, we can still obtain nearly constant coupling efficiency to the TEM$_{00}$ mode of an external femtosecond enhancement cavity (fsEC) for the full power range of the laser by simply changing the lens spacing in a mode-matching telescope between the compressor and the fsEC. One of the authors (T. K. Allison) has observed similar behavior for the laser described in reference \citenum{Ruehl_OptLett2010}.

The dispersion budget for the 80 W comb is shown in table \ref{table:80Wtable} and the raw FROG trace and retrieved pulse shape are shown in figure \ref{fig:80Wresult}c). The compressor has an overall efficiency of 86\%, corresponding to a grating diffraction efficiency of $(0.86)^{1/4} = 96\%$. Clean 155 fs pulses are observed at both high and low power. The absence of pulse distortion at high power indicates that linear amplification has been achieved.

\begin{figure}[t]
    \centering
   \includegraphics[width=\linewidth]{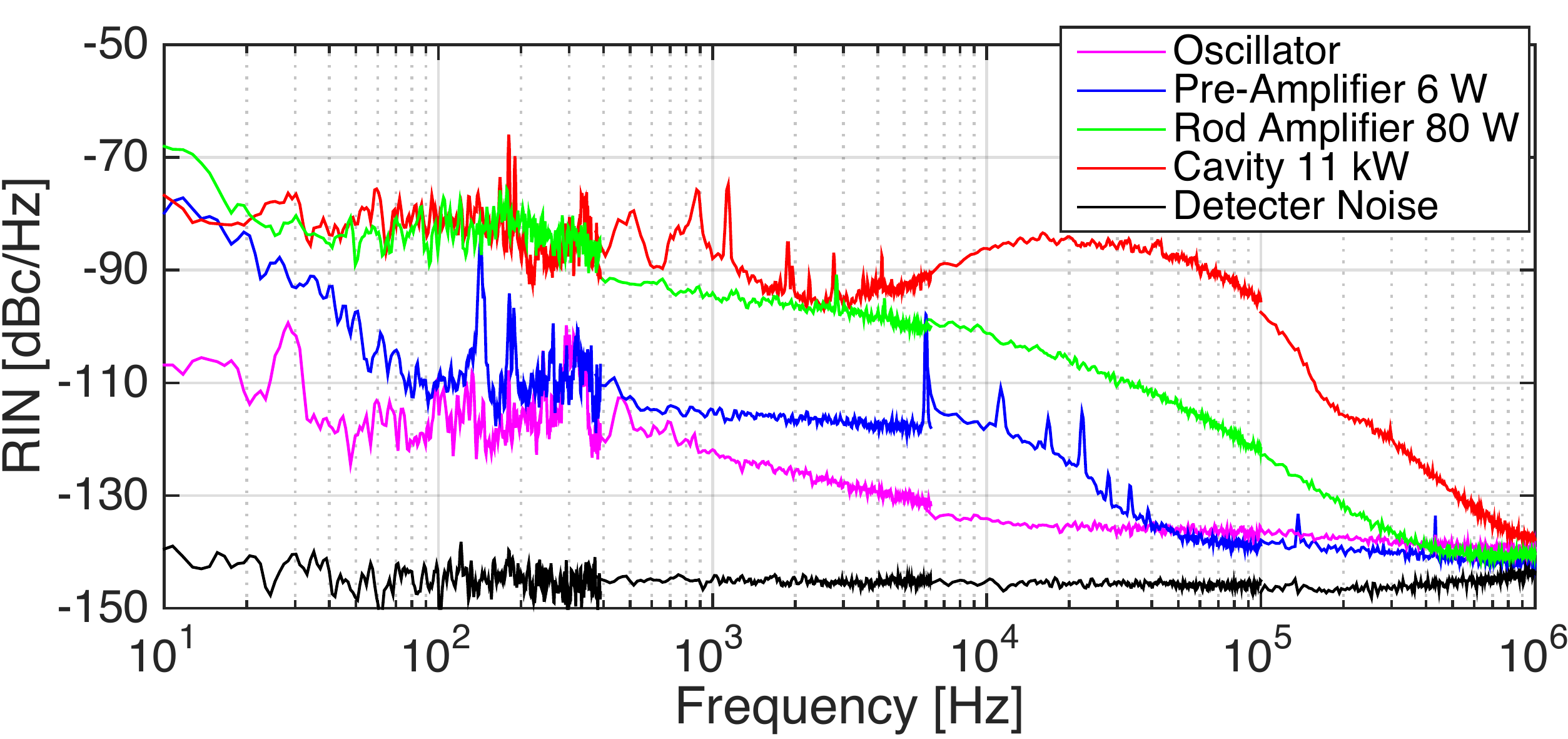}
    \caption{\small \textbf{Intensity noise.}  RIN spectra at various places throughout the 80 W laser system. The RIN spectrum of the 9 W laser is similar to the blue curve.}
    \label{fig:RIN}
\end{figure}

Figure \ref{fig:RIN} shows the RIN measured in various parts of the laser system. The pump diode RIN spectra are nearly flat out to frequencies of 1 MHz at -100 dBc/Hz for the 30 W pump diode and -85 dBc/Hz for the 200 W pump diode. The 200 W pump diode is driven by a 1500 W power supply from TDK Lambda (GEN 20-76). For the both the pre-amplifier and the rod amplifier, the storage time of the Yb:fiber gain medium provides a low-pass filter to the pump diode RIN. Figure \ref{fig:RIN} also shows the RIN spectrum for the intracavity light of a fsEC operating with 11 kW of intracavity power. The cavity is locked to the comb using the PDH method\cite{Black_AmJPhys2001,Drever_APB1983} and a PZT on one of the fsEC mirrors. A servo bump is observed at $\sim$30 kHz due to the finite bandwidth of the PZT lock, but for lower frequencies an intracavity RIN level is obtained nearly equal to the RIN of the comb. Lower RIN for a high-power Yb:fiber laser has been obtained recently by Wunram \emph{et al.},\cite{Wunram_OptLett2015} however the dominant noise source was not identified in this work.





\section{Applications}\label{sec:apps}

In our laboratory, the primary application of these frequency combs involves coupling them to passive optical cavities\cite{Jones_OptLett2004, Gherman_OptExp2002} for either performing spectroscopy\cite{Reber_Optica2016} or generating high-order harmonics.\cite{Corder_DAMOP2016} Here we give examples using these lasers in both applications. 

\subsection{Ultrasensitive femtosecond optical spectroscopy}

Ultrafast optical spectroscopy methods, such as transient absorption spectroscopy and 2D spectroscopy, are widely used across many disciplines. However, these techniques are typically restricted to optically thick samples, such as solids and liquid solutions, since the sensitivity of femtosecond-time resolved optical spectroscopy has lagged far behind that of linear spectroscopy methods.\cite{Davis_ChemPhysLett2001, Ye_JOSAB1998, Gagliardi_Book2013} 

Recently we have demonstrated the cavity-enhancement of femtosecond time-resolved measurements using the 9 W comb described here, enabling ultrafast optical spectroscopy in dilute molecular beams.\cite{Reber_Optica2016} Complete details of the optical setup are given in reference \citenum{Reber_Optica2016} but here we give a brief summary. Light from the 9 W comb is frequency doubled and sent to two 4-mirror ring cavities  whose foci intersect in a supersonic expansion of I$_2$ seeded in a noble carrier gas. A two point locking scheme is used to stabilize the laser to the probe cavity. The center of the comb is tightly locked to the probe fsEC using the PDH method\cite{Jones_OptLett2004, Black_AmJPhys2001,Drever_APB1983} and the long EOM in oscillator A. A bias tee is used to provide both PDH FM sidebands and a high voltage feedback signal to the EOM. Using a wideband high voltage amplifier (Thorlabs inc. HVA200), we achieve approximately 100 kHz bandwidth in this feedback loop. A second, slower, PDH feedback loop monitors a separate part of the optical spectrum and moves the intracavity grating separation to maintain resonance for the rest of the comb.

\begin{figure}[t]
    \centering
   \includegraphics[width=\linewidth]{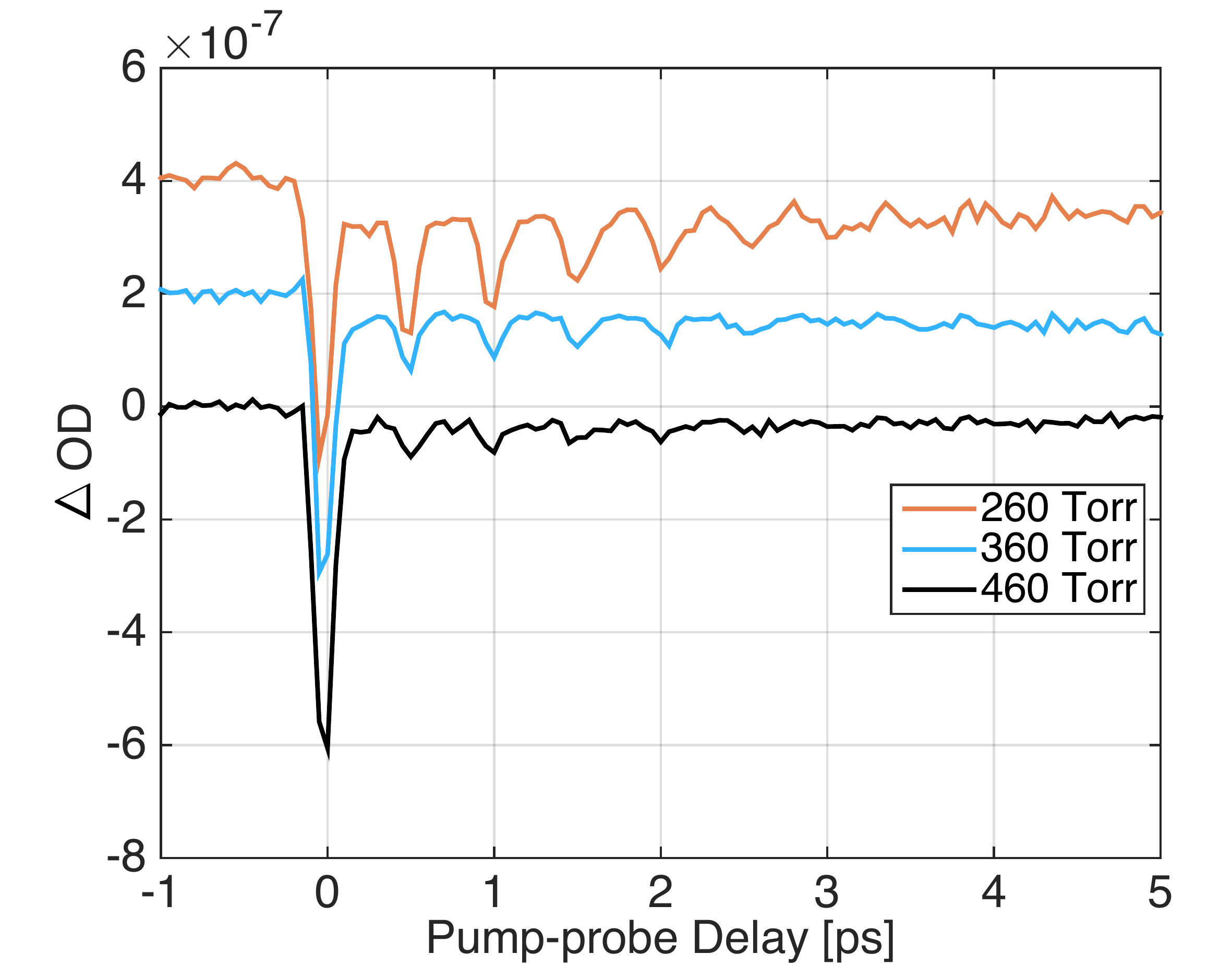}
    \caption{\small \textbf{Ultrasensitive transient absorption spectroscopy}. The comb from the 1060 nm laser is frequency doubled and coupled to two femtosecond enhancement cavities for performing ultrafast spectroscopy, as in ref. \citenum{Reber_Optica2016}. Shown are pump-probe traces in a supersonic expansion of molecular iodine and Argon gas, 6 mm above a 200 $\mu$m wide slit nozzle. As the Argon pressure is increased, Argon atoms cluster around the I$_2$ molecules, quenching the excited state vibrational wavepacket. Curves are offset for clarity.}
    \label{fig:I2Ar}
\end{figure}

In reference \citenum{Reber_Optica2016} we reported measurements taken in a helium-seeded supersonic expansion using a 700 $\mu$m diameter round nozzle, where there is no clustering between the He and the I$_2$ molecules. If the expansion is instead seeded with Argon under the right conditions, the effects of clustering can be observed as the backing pressure is increased. Figure \ref{fig:I2Ar} shows transient absorption data taken for I$_2$ in an Argon-seeded expansion, 6 mm away from a 5 mm long $\times$ 200 $\mu$m wide slit nozzle  at three different stagnation pressures. The nozzle is 3D printed in PEEK plastic (Arevo Labs Inc.) for compatibility with I$_2$. As the backing pressure is increased from 260 Torr to 460 Torr (absolute), the effects of clustering are clearly seen as the coherent transient peak at time zero increases and subsequent oscillations are suppressed as the I$_2$ molecules collide with the caging Argon atoms.\cite{Wang_JPhysChem1995}

\subsection{Cavity-enhanced high-order harmonic generation and extreme-UV frequency combs.}

High-power ultrafast lasers are now routinely used to drive high harmonic generation (HHG), providing extreme ultraviolet (XUV) ultrashort pulses with a table-top setup, and also attosecond pulses.\cite{Brabec_RMP2000, Chini_nph2014} This is a highly nonlinear process, during which electrons are ionized by a strong laser field and then recombine  with their parent ions, generating coherent light at high photon energies. Conventionally, HHG is realized by focusing high energy ($>$ 100 $\mu$J) laser pulses to intensities of more than 10$^{13}$ W/cm$^2$ intensity in a noble gas. The repetition rate is then limited by the average laser power available, usually to less than 100 kHz, but many applications such as XUV frequency metrology,\cite{Cingoz_Nature2012} surface photoemission,\cite{Passlack_JAP2006} and photoionization coincidence methods\cite{Sandhu_Science2008} demand higher repetition rates.
\begin{figure}[t]
    \centering
   \includegraphics[width=\linewidth]{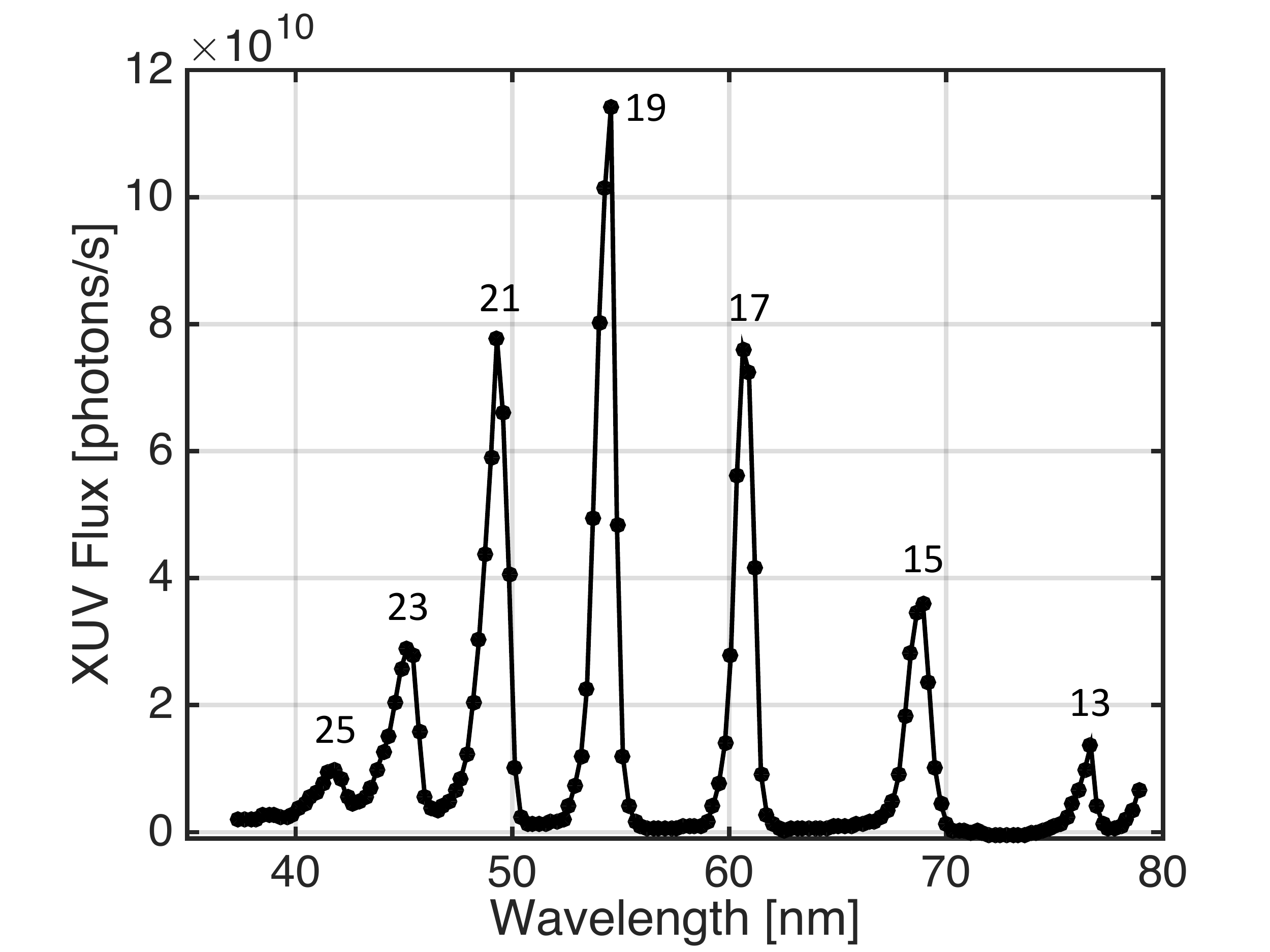}
    \caption{\small \textbf{Cavity-enhanced high-harmonic generation}. Spectrum of harmonics generated from Krypton gas via cavity-enhanced HHG at 87 MHz repetition rate. Photon fluxes correspond to values measured at the exit slit of a time-preserving monochromator, corresponding to the flux directly deliverable to experiments.}
    \label{fig:HHG}
\end{figure}

An elegant way to achieve this is via resonant enhancement of the fundamental pulse train in a fsEC.\cite{Jones_OptLett2002, Jones_OptLett2004} Although generating high harmonics in these cavities was originally demonstrated as early as 2005,\cite{Gohle_Nature2005, Jones_PRL2005} since then through the use of higher power driving lasers and understanding of the intracavity extreme-nonlinear optics\cite{Allison_PRL2011, Carlson_OptLett2011, Yost_OptExp2011, Holzberger_PRL2015} the power from cavity-enhanced HHG systems has increased by more than 6 orders of magnitude, to more the 100 $\mu$W/harmonic (at the gas jet) for ~20 eV harmonics generated in Xenon.\cite{Cingoz_Nature2012,Allison_PRL2011,Yost_OptExp2011,Pupeza_PRL2014, Lee_OptExp2011}

In our lab, we have built a cavity-enhanced HHG system for the purpose of performing time-resolved surface photoemission experiments at high repetition rate. The 1035 nm comb described in this paper is coupled to a 6 mirror femtosecond enhancement cavity with a 1\% input coupler and roughly 15 $\mu$m focus size. Krypton gas is injected at the focus using a small glass capillary backed with 760 Torr pressure and harmonics are coupled out of the cavity using a 250 $\mu$m thick sapphire plate at Brewster's angle for the fundamental. The harmonics are then passed through a home-built time-preserving monochromator similar to that described by Frasetto et al.\cite{Frassetto_OptExp2011} Figure \ref{fig:HHG} shows the spectrum of harmonics after the monochromator exit slit measured using an Al-coated silicon photodiode (Opto Diode AXUV100Al) as the grating angle is rocked. For calculating the y-axis, the manufacturer's quantum efficiency specifications were used without additional calibration.  More than $10^{11}$ photons per second emerge from the monochromator in the 19th harmonic at 23 eV, or approximately 1400 photons/pulse. The low pulse energy and high repetition rate make this source an ideal system for time-resolved photoelectron spectroscopy or time-resolved photoelectron microscopy experiments which are limited by space charge effects. \cite{Chew_BookChapter2015, Plotzing_RSI2016}

\section*{Acknowledgements}

This material is based upon work supported by the Air Force Office of Scientific Research under award numbers FAFA9550-16-1-0164 and FA9550-13-1-0109, and the National Science Foundation under award number 1404296. When we embarked on this laser building project three years ago, we were helped immensely by S. A. Diddams, G. Ycas, and L. Nugent-Glandorf at the National Institute of Standards and Technology in Boulder, CO and D. J. Jones and A. K. Mills at the University of British Columbia. We also thank R. Jason Jones at the University of Arizona for useful discussions.

\small

\begin{thebibliography}{100}

\bibitem{Newbury_NatPhot2011}
Nathan~R. Newbury.
\newblock Searching for applications with a fine-tooth comb.
\newblock {\em Nat Photon}, 5(4):186--188, 04 2011.

\bibitem{Diddams_JOSAB2010}
Scott~A. Diddams.
\newblock The evolving optical frequency comb.
\newblock {\em J. Opt. Soc. Am. B}, 27(11):B51--B62, Nov 2010.

\bibitem{Steinmetz_Science2008}
Tilo Steinmetz, Tobias Wilken, Constanza Araujo-Hauck, Ronald Holzwarth,
  Theodor~W. H{\"a}nsch, Luca Pasquini, Antonio Manescau, Sandro
  D{\textquoteright}Odorico, Michael~T. Murphy, Thomas Kentischer, Wolfgang
  Schmidt, and Thomas Udem.
\newblock Laser frequency combs for astronomical observations.
\newblock {\em Science}, 321(5894):1335--1337, 2008.

\bibitem{Coddington_NatPhot2009}
CoddingtonI., SwannW. C., NenadovicL., and NewburyN. R.
\newblock Rapid and precise absolute distance measurements at long range.
\newblock {\em Nat Photon}, 3(6):351--356, 06 2009.

\bibitem{Mills_JPhysB2012}
Arthur~K Mills, T~J Hammond, Matthew H~C Lam, and David~J Jones.
\newblock Xuv frequency combs via femtosecond enhancement cavities.
\newblock {\em Journal of Physics B: Atomic, Molecular and Optical Physics},
  45(14):142001, 2012.

\bibitem{Krausz_RMP2009}
Ferenc Krausz and Misha Ivanov.
\newblock Attosecond physics.
\newblock {\em Rev. Mod. Phys.}, 81(1):163--234, Feb 2009.

\bibitem{Benko_NatPhot2014}
Craig Benko, Thomas~K. Allison, Arman Cingoz, Linqiang Hua, Francois Labaye,
  Dylan~C. Yost, and Jun Ye.
\newblock Extreme ultraviolet radiation with coherence time greater than 1 s.
\newblock {\em Nat Photon}, 8(7):530--536, 07 2014.

\bibitem{Adler_AnnRevChem2010}
Florian Adler, Michael~J. Thorpe, Kevin~C. Cossel, and Jun Ye.
\newblock Cavity-enhanced direct frequency comb spectroscopy: Technology and
  applications.
\newblock {\em Annual Review of Analytical Chemistry}, 3(1):175--205, 2010.

\bibitem{Coddington_Optica2016}
Ian Coddington, Nathan Newbury, and William Swann.
\newblock Dual-comb spectroscopy.
\newblock {\em Optica}, 3(4):414--426, Apr 2016.

\bibitem{Reber_Optica2016}
Melanie A.~R. Reber, Yuning Chen, and Thomas~K. Allison.
\newblock Cavity-enhanced ultrafast spectroscopy: ultrafast meets
  ultrasensitive.
\newblock {\em Optica}, 3(3):311--317, Mar 2016.

\bibitem{Adler_OptLett2009}
Florian Adler, Kevin~C. Cossel, Michael~J. Thorpe, Ingmar Hartl, Martin~E.
  Fermann, and Jun Ye.
\newblock Phase-stabilized, 1.5 w frequency comb at 2.8--4.8 $\mu$m.
\newblock {\em Opt. Lett.}, 34(9):1330--1332, May 2009.

\bibitem{Cingoz_Nature2012}
Arman Cing\"{o}z, Dylan~C. Yost, Thomas~K. Allison, Axel Ruehl, Martin~E.
  Fermann, Ingmar Hartl, and Jun Ye.
\newblock Direct frequency comb spectroscopy in the extreme ultraviolet.
\newblock {\em Nature}, 482(7383):68--71, 02 2012.

\bibitem{LeinDecker_OptExp2012}
Nick Leindecker, Alireza Marandi, Robert~L. Byer, Konstantin~L. Vodopyanov, Jie
  Jiang, Ingmar Hartl, Martin Fermann, and Peter~G. Schunemann.
\newblock Octave-spanning ultrafast opo with 2.6-6.1 micron instantaneous
  bandwidth pumped by femtosecond tm-fiber laser.
\newblock {\em Opt. Express}, 20(7):7046--7053, Mar 2012.

\bibitem{Cruz_OptExp2015}
Flavio~C. Cruz, Daniel~L. Maser, Todd Johnson, Gabriel Ycas, Andrew Klose,
  Fabrizio~R. Giorgetta, Ian Coddington, and Scott~A. Diddams.
\newblock Mid-infrared optical frequency combs based on difference frequency
  generation for molecular spectroscopy.
\newblock {\em Opt. Express}, 23(20):26814--26824, Oct 2015.

\bibitem{Paschotta_IEEE1997}
R.~Paschotta, J.~Nilsson, A.C. Tropper, and D.C. Hanna.
\newblock Ytterbium-doped fiber amplifiers.
\newblock {\em Quantum Electronics, IEEE Journal of}, 33(7):1049--1056, 1997.

\bibitem{Fermann_IEEE2009}
M.E. Fermann and I.~Hartl.
\newblock Ultrafast fiber laser technology.
\newblock {\em Selected Topics in Quantum Electronics, IEEE Journal of},
  15(1):191 --206, jan. 2009.

\bibitem{Fattahi_Optica2014}
Hanieh Fattahi, Helena~G. Barros, Martin Gorjan, Thomas Nubbemeyer, Bidoor
  Alsaif, Catherine~Y. Teisset, Marcel Schultze, Stephan Prinz, Matthias
  Haefner, Moritz Ueffing, Ayman Alismail, L\'{e}n\'{a}rd V\'{a}mos, Alexander
  Schwarz, Oleg Pronin, Jonathan Brons, Xiao~Tao Geng, Gunnar Arisholm, Marcelo
  Ciappina, Vladislav~S. Yakovlev, Dong-Eon Kim, Abdallah~M. Azzeer, Nicholas
  Karpowicz, Dirk Sutter, Zsuzsanna Major, Thomas Metzger, and Ferenc Krausz.
\newblock Third-generation femtosecond technology.
\newblock {\em Optica}, 1(1):45--63, Jul 2014.

\bibitem{Cautaerts_OptLett1997}
V.~Cautaerts, D.~J. Richardson, R.~Paschotta, and D.~C. Hanna.
\newblock Stretched pulse yb3$+$:silica fiber laser.
\newblock {\em Opt. Lett.}, 22(5):316--318, Mar 1997.

\bibitem{Honninger_OptLett1995}
C.~H\"{o}nninger, A.~Giesen, G.~Zhang, and U.~Keller.
\newblock Femtosecond yb:yag laser using semiconductor saturable absorbers.
\newblock {\em Opt. Lett.}, 20(23):2402--2404, Dec 1995.

\bibitem{Eidam_OptLett2010}
Tino Eidam, Stefan Hanf, Enrico Seise, Thomas~V. Andersen, Thomas Gabler,
  Christian Wirth, Thomas Schreiber, Jens Limpert, and Andreas T\"{u}nnermann.
\newblock Femtosecond fiber cpa system emitting 830 w average output power.
\newblock {\em Opt. Lett.}, 35(2):94--96, Jan 2010.

\bibitem{Jauregui_nph2013}
Cesar Jauregui, Jens Limpert, and Andreas Tunnermann.
\newblock High-power fibre lasers.
\newblock {\em Nat Photon}, 7(11):861--867, 11 2013.

\bibitem{Schibli_NatPhot2008}
Schibli~T. R., Hartl I., Yost~D. C., Martin~M. J., Marcinkevicius A.,
  Fermann~M. E., and Ye~J.
\newblock Optical frequency comb with submillihertz linewidth and more than 10
  w average power.
\newblock {\em Nat Photon}, 2(6):355--359, 06 2008.

\bibitem{Kuznetsova_APB2007}
L.~Kuznetsova, F.W. Wise, S.~Kane, and J.~Squier.
\newblock Chirped-pulse amplification near the gain-narrowing limit of yb-doped
  fiber using a reflection grism compressor.
\newblock {\em Applied Physics B}, 88(4):515--518, 2007.

\bibitem{Zhao_OptExp2014}
Jian Zhao, Wenxue Li, Chao Wang, Yang Liu, and Heping Zeng.
\newblock Pre-chirping management of a self-similar yb-fiber amplifier towards
  80 w average power with sub-40 fs pulse generation.
\newblock {\em Opt. Express}, 22(26):32214--32219, Dec 2014.

\bibitem{Liu_OptLett2015}
Wei Liu, Damian~N. Schimpf, Tino Eidam, Jens Limpert, Andreas T\"{u}nnermann,
  Franz~X. K\"{a}rtner, and Guoqing Chang.
\newblock Pre-chirp managed nonlinear amplification in fibers delivering 100 w,
  60 fs pulses.
\newblock {\em Opt. Lett.}, 40(2):151--154, Jan 2015.

\bibitem{Hartl_OptLett2007}
I.~Hartl, T.~R. Schibli, A.~Marcinkevicius, D.~C. Yost, D.~D. Hudson, M.~E.
  Fermann, and Jun Ye.
\newblock Cavity-enhanced similariton yb-fiber laser frequency comb: $3 \times
  10^{14}$ w/cm$^2$ peak intensity at 136 mhz.
\newblock {\em Opt. Lett.}, 32(19):2870--2872, 2007.

\bibitem{Ruehl_OptLett2010}
Axel Ruehl, Andrius Marcinkevicius, Martin~E. Fermann, and Ingmar Hartl.
\newblock 80 w, 120 fs yb-fiber frequency comb.
\newblock {\em Opt. Lett.}, 35(18):3015--3017, Sep 2010.

\bibitem{Nugent-Glandorf_OptLett2011}
Lora Nugent-Glandorf, Todd~A. Johnson, Yohei Kobayashi, and Scott~A. Diddams.
\newblock Impact of dispersion on amplitude and frequency noise in a yb-fiber
  laser comb.
\newblock {\em Opt. Lett.}, 36(9):1578--1580, May 2011.

\bibitem{Jones_OptLett2002}
R.~Jason Jones and Jun Ye.
\newblock Femtosecond pulse amplification by coherent addition in a passive
  optical cavity.
\newblock {\em Opt. Lett.}, 27(20):1848--1850, 2002.

\bibitem{Jones_OptLett2004}
R.~Jason Jones and Jun Ye.
\newblock High-repetition-rate coherent femtosecond pulse amplification withan
  external passive optical cavity.
\newblock {\em Opt. Lett.}, 29(23):2812--2814, Dec 2004.

\bibitem{Carstens_OptLett2014}
H.~Carstens, N.~Lilienfein, S.~Holzberger, C.~Jocher, T.~Eidam, J.~Limpert,
  A.~T\"{u}nnermann, J.~Weitenberg, D.~C. Yost, A.~Alghamdi, Z.~Alahmed,
  A.~Azzeer, A.~Apolonski, E.~Fill, F.~Krausz, and I.~Pupeza.
\newblock Megawatt-scale average-power ultrashort pulses in an enhancement
  cavity.
\newblock {\em Opt. Lett.}, 39(9):2595--2598, May 2014.

\bibitem{Paul_OptLett2008}
Justin Paul, James Johnson, Jane Lee, and R.~Jason Jones.
\newblock Generation of high-power frequency combs from injection-locked
  femtosecond amplification cavities.
\newblock {\em Opt. Lett.}, 33(21):2482--2484, Nov 2008.

\bibitem{Chong_RepProgPhys2015}
Andy Chong, Logan~G Wright, and Frank~W Wise.
\newblock Ultrafast fiber lasers based on self-similar pulse evolution: a
  review of current progress.
\newblock {\em Reports on Progress in Physics}, 78(11):113901, 2015.

\bibitem{Fernandez_OptLett2012}
A.~Fern\'{a}ndez, K.~Jespersen, L.~Zhu, L.~Gr\"{u}ner-Nielsen, A.~Baltu\v{s}ka,
  A.~Galvanauskas, and A.~J. Verhoef.
\newblock High-fidelity, 160�fs, 5�$\mu$j pulses from an integrated
  yb-fiber laser system with a fiber stretcher matching a simple grating
  compressor.
\newblock {\em Opt. Lett.}, 37(5):927--929, Mar 2012.

\bibitem{Dudley_NatPhys2007}
John~M. Dudley, Christophe Finot, David~J. Richardson, and Guy Millot.
\newblock Self-similarity in ultrafast nonlinear optics.
\newblock {\em Nat Phys}, 3(9):597--603, 09 2007.

\bibitem{Fermann_PRL2000}
M.~E. Fermann, V.~I. Kruglov, B.~C. Thomsen, J.~M. Dudley, and J.~D. Harvey.
\newblock Self-similar propagation and amplification of parabolic pulses in
  optical fibers.
\newblock {\em Phys. Rev. Lett.}, 84:6010--6013, Jun 2000.

\bibitem{Agrawal_AppNonlinearFiberOptics}
G.~P. Agrawal.
\newblock {\em Applications of Nonlinear Fiber Optics}.
\newblock Academic Press, 2008.

\bibitem{Agrawal_NonlinearFiberOpticsBook}
G.~P. Agrawal.
\newblock {\em Nonlinear Fiber Optics}.
\newblock Academic Press, 2012.

\bibitem{Asaki_OptLett1993}
Melanie~T. Asaki, Chung-Po Huang, Dennis Garvey, Jianping Zhou, Henry~C.
  Kapteyn, and Margaret~M. Murnane.
\newblock Generation of 11-fs pulses from a self-mode-locked ti:sapphire laser.
\newblock {\em Opt. Lett.}, 18(12):977--979, 1993.

\bibitem{Stingl_OptLett1995}
A.~Stingl, R.~Szip\"{o}cs, M.~Lenzner, Ch. Spielmann, and F.~Krausz.
\newblock Sub-10-fs mirror-dispersion-controlled ti:sapphire laser.
\newblock {\em Opt. Lett.}, 20(6):602--604, Mar 1995.

\bibitem{Kobtsev_OptExp2008}
Sergey Kobtsev, Sergey Kukarin, and Yurii Fedotov.
\newblock Ultra-low repetition rate mode-locked fiber laser with high-energy
  pulses.
\newblock {\em Opt. Express}, 16(26):21936--21941, Dec 2008.

\bibitem{Usechak_OptLett2004}
Nick~G. Usechak, Govind~P. Agrawal, and Jonathan~D. Zuegel.
\newblock Tunable, high-repetition-rate, harmonically mode-locked ytterbium
  fiber laser.
\newblock {\em Opt. Lett.}, 29(12):1360--1362, Jun 2004.

\bibitem{Chong_OptExp2006}
Andy Chong, Joel Buckley, Will Renninger, and Frank Wise.
\newblock All-normal-dispersion femtosecond fiber laser.
\newblock {\em Opt. Express}, 14(21):10095--10100, Oct 2006.

\bibitem{Kelly_ElecLetters1992}
S.~M.~J. Kelly.
\newblock Characteristic sideband instability of periodically amplified average
  soliton.
\newblock {\em Electronics Letters}, 28(8):806--807, April 1992.

\bibitem{BaumGartl_OptLett2012}
Martin Baumgartl, Caroline Lecaplain, Ammar Hideur, Jens Limpert, and Andreas
  T\"{u}nnermann.
\newblock 66�w average power from a microjoule-class sub-100�fs fiber
  oscillator.
\newblock {\em Opt. Lett.}, 37(10):1640--1642, May 2012.

\bibitem{Paschotta_ApplPhysB2004_1}
R.~Paschotta.
\newblock Noise of mode-locked lasers (part i): numerical model.
\newblock {\em Applied Physics B}, 79(2):153--162, 2004.

\bibitem{Paschotta_ApplPhysB2004_2}
R.~Paschotta.
\newblock Noise of mode-locked lasers (part ii): timing jitter and other
  fluctuations.
\newblock {\em Applied Physics B}, 79(2):163--173, 2004.

\bibitem{Hartl_CLEO2005}
Ingmar Hartl, Genady Imeshev, Liang Dong, Gyu~C. Cho, and Martin~E. Fermann.
\newblock Ultra-compact dispersion compensated femtosecond fiber oscillators
  and amplifiers.
\newblock In {\em Conference on Lasers and Electro-Optics/Quantum Electronics
  and Laser Science and Photonic Applications Systems Technologies}, page
  CThG1. Optical Society of America, 2005.

\bibitem{Zhou_OptExp2008}
Xiangyu Zhou, Dai Yoshitomi, Yohei Kobayashi, and Kenji Torizuka.
\newblock Generation of 28-fs pulses from a mode-locked ytterbium fiber
  oscillator.
\newblock {\em Opt. Express}, 16(10):7055--7059, May 2008.

\bibitem{Buckley_OptLett2006}
J.~R. Buckley, S.~W. Clark, and F.~W. Wise.
\newblock Generation of ten-cycle pulses from an ytterbium fiber laser with
  cubic phase compensation.
\newblock {\em Opt. Lett.}, 31(9):1340--1342, May 2006.

\bibitem{Hofer_IEEEJQE1992}
M.~Hofer, M.H. Ober, F.~Haberl, and M.E. Fermann.
\newblock Characterization of ultrashort pulse formation in passively
  mode-locked fiber lasers.
\newblock {\em Quantum Electronics, IEEE Journal of}, 28(3):720--728, 1992.

\bibitem{Weiner_book2009}
Andrew Weiner.
\newblock {\em Ultrafast Optics}.
\newblock Wiley, 2009.

\bibitem{Newbury_talk}
N.~Newbury.
\newblock Understanding noise sources and stabilization strategies in frequency
  combs - part 2.
\newblock Winter College on Optics: Optical Frequency Combs, 2016.

\bibitem{Sinclair_RSI2015}
L.~C. Sinclair, J.-D. Desch√{\texttrademark}nes, L.~Sonderhouse, W.~C. Swann,
  I.~H. Khader, E.~Baumann, N.~R. Newbury, and I.~Coddington.
\newblock Invited article: A compact optically coherent fiber frequency comb.
\newblock {\em Review of Scientific Instruments}, 86(8), 2015.

\bibitem{Note1}
The quiet Yb-oscillators described in references\protect \citenum
  {Schibli_NatPhot2008} and \protect \citenum {Ruehl_OptLett2010} actually use
  a combination of both NPE and the SESAM.

\bibitem{Ilday_OE2005}
F.~\"{O} Ilday, J.~Chen, and F.~X. K\"{a}rtner.
\newblock Generation of sub-100-fs pulses at up to 200 mhz repetition rate from
  a passively mode-locked yb-doped fiber laser.
\newblock {\em Opt. Express}, 13(7):2716--2721, 2005.

\bibitem{Newbury_JOSAB2007}
Nathan~R. Newbury and William~C. Swann.
\newblock Low-noise fiber-laser frequency combs (invited).
\newblock {\em J. Opt. Soc. Am. B}, 24(8):1756--1770, 2007.

\bibitem{Bock_thesis}
Katherine~J. Bock.
\newblock {\em Femtosecond Fiber Lasers}.
\newblock PhD thesis, School of Electrical Engineering and Computer Science,
  University of Ottawa, Ottawa, Canada, 2012.

\bibitem{Ranka_OptLett1997}
Jinendra~K. Ranka, Alexander~L. Gaeta, Andrius Baltuska, Maxim~S.
  Pshenichnikov, and Douwe~A. Wiersma.
\newblock Autocorrelation measurement of 6-fs pulses based on the
  two-photon-induced photocurrent in a gaasp photodiode.
\newblock {\em Opt. Lett.}, 22(17):1344--1346, Sep 1997.

\bibitem{Knox_OL1992}
W.~H. Knox.
\newblock In situ measurement of complete intracavitydispersion in an operating
  ti:sapphire femtosecond laser.
\newblock {\em Opt. Lett.}, 17(7):514--516, 1992.

\bibitem{Cingoz_OptLett2011}
A.~Cing\"{o}z, D.~C. Yost, T.~K. Allison, A.~Ruehl, M.~E. Fermann, I.~Hartl,
  and J.~Ye.
\newblock Broadband phase noise suppression in a yb-fiber frequency comb.
\newblock {\em Opt. Lett.}, 36(5):743--745, Mar 2011.

\bibitem{Song_OE2011}
Youjian Song, Chur Kim, Kwangyun Jung, Hyoji Kim, and Jungwon Kim.
\newblock Timing jitter optimization of mode-locked yb-fiber lasers toward the
  attosecond regime.
\newblock {\em Opt. Express}, 19(15):14518--14525, 2011.

\bibitem{Udem_ICTP2016}
Udem Th.
\newblock The frequency comb (r)evolution.
\newblock In {\em Winter College on Optics: Optical Frequency Combs - from
  multispecies gas sensing to high precision interrogation of atomic and
  molecular targets}, International Centre for Theoretical Physics, Feb. 2016.

\bibitem{Udem_OptLett1999}
Th. Udem, J.~Reichert, R.~Holzwarth, and T.~W. H\"{a}nsch.
\newblock Accurate measurement of large optical frequency differences with a
  mode-locked laser.
\newblock {\em Opt. Lett.}, 24(13):881--883, Jul 1999.

\bibitem{Saha_OptExp2013}
Kasturi Saha, Yoshitomo Okawachi, Bonggu Shim, Jacob~S. Levy, Reza Salem,
  Adrea~R. Johnson, Mark~A. Foster, Michael R.~E. Lamont, Michal Lipson, and
  Alexander~L. Gaeta.
\newblock Modelocking and femtosecond pulse generation in chip-based frequency
  combs.
\newblock {\em Opt. Express}, 21(1):1335--1343, Jan 2013.

\bibitem{Kourogi_IEEE1993}
M.~Kourogi, K.~Nakagawa, and M.~Ohtsu.
\newblock Wide-span optical frequency comb generator for accurate optical
  frequency difference measurement.
\newblock {\em IEEE Journal of Quantum Electronics}, 29(10):2693--2701, Oct
  1993.

\bibitem{Kuse_OE2015}
N.~Kuse, C.-C. Lee, J.~Jiang, C.~Mohr, T.~R. Schibli, and M.E. Fermann.
\newblock Ultra-low noise all polarization-maintaining er fiber-based optical
  frequency combs facilitated with a graphene modulator.
\newblock {\em Opt. Express}, 23(19):24342--24350, Sep 2015.

\bibitem{Hudson_OL2005}
Darren~D. Hudson, Kevin~W. Holman, R.~Jason Jones, Steven~T. Cundiff, Jun Ye,
  and David~J. Jones.
\newblock Mode-locked fiber laser frequency-controlled with an intracavity
  electro-optic modulator.
\newblock {\em Opt. Lett.}, 30(21):2948--2950, 2005.

\bibitem{Zhang_IEEE2012}
W.~Zhang, M.~Lours, M.~Fischer, R.~Holzwarth, G.~Santarelli, and Y.~L. Coq.
\newblock Characterizing a fiber-based frequency comb with electro-optic
  modulator.
\newblock {\em IEEE Transactions on Ultrasonics, Ferroelectrics, and Frequency
  Control}, 59(3):432--438, 2012.

\bibitem{Iwakuni_OE2012}
Kana Iwakuni, Hajime Inaba, Yoshiaki Nakajima, Takumi Kobayashi, Kazumoto
  Hosaka, Atsushi Onae, and Feng-Lei Hong.
\newblock Narrow linewidth comb realized with a mode-locked fiber laser using
  an intra-cavity waveguide electro-optic modulator for high-speed control.
\newblock {\em Opt. Express}, 20(13):13769--13776, 2012.

\bibitem{Swann_OE2011}
William~C. Swann, Esther Baumann, Fabrizio~R. Giorgetta, and Nathan~R. Newbury.
\newblock Microwave generation with low residual phase noise from a femtosecond
  fiber laser with an intracavity electro-optic modulator.
\newblock {\em Opt. Express}, 19(24):24387--24395, Nov 2011.

\bibitem{Nakajima_OE2010}
Yoshiaki Nakajima, Hajime Inaba, Kazumoto Hosaka, Kaoru Minoshima, Atsushi
  Onae, Masami Yasuda, Takuya Kohno, Sakae Kawato, Takao Kobayashi, Toshio
  Katsuyama, and Feng-Lei Hong.
\newblock A multi-branch, fiber-based frequency comb with millihertz-level
  relative linewidths using an intra-cavity electro-optic modulator.
\newblock {\em Opt. Express}, 18(2):1667--1676, Jan 2010.

\bibitem{Benko_OptLett2012}
C.~Benko, A.~Ruehl, M.~J. Martin, K.~S.~E. Eikema, M.~E. Fermann, I.~Hartl, and
  J.~Ye.
\newblock Full phase stabilization of a yb:fiber femtosecond frequency comb via
  high-bandwidth transducers.
\newblock {\em Opt. Lett.}, 37(12):2196--2198, Jun 2012.

\bibitem{Lee_OL2012}
C.-C. Lee, C.~Mohr, J.~Bethge, S.~Suzuki, M.~E. Fermann, I.~Hartl, and T.~R.
  Schibli.
\newblock Frequency comb stabilization with bandwidth beyond the limit of gain
  lifetime by an intracavity graphene electro-optic modulator.
\newblock {\em Opt. Lett.}, 37(15):3084--3086, 2012.

\bibitem{Hellwig_OL2014}
Tim Hellwig, Steffen Rieger, and Carsten Fallnich.
\newblock Toward an all-optically stabilized frequency comb based on a
  mode-locked fiber laser.
\newblock {\em Opt. Lett.}, 39(3):525--527, 2014.

\bibitem{Bao_Ol2014}
Chengying Bao, Andrew~C. Funk, Changxi Yang, and Steven~T. Cundiff.
\newblock Pulse dynamics in a mode-locked fiber laser and its quantum limited
  comb frequency uncertainty.
\newblock {\em Opt. Lett.}, 39(11):3266--3269, 2014.

\bibitem{Kuse_OL2010}
Naoya Kuse, Yutaka Nomura, Akira Ozawa, Makoto Kuwata-Gonokami, Shuntaro
  Watanabe, and Yohei Kobayashi.
\newblock Self-compensation of third-order dispersion for ultrashort pulse
  generation demonstrated in an yb fiber oscillator.
\newblock {\em Opt. Lett.}, 35(23):3868--3870, 2010.

\bibitem{Hyodo_OptComm1999}
Masaharu Hyodo, Kazi~Sarwar Abedin, and Noriaki Onodera.
\newblock Generation of millimeter-wave signals up to 70.5 ghz by heterodyning
  of two extended-cavity semiconductor lasers with an intracavity electro-optic
  crystal.
\newblock {\em Optics Communications}, 171(1--3):159 -- 169, 1999.

\bibitem{Black_AmJPhys2001}
E.~D. Black.
\newblock An introduction to pound-drever-hall laser stabilization.
\newblock {\em Am. J. Phys.}, page~79, 2001.

\bibitem{Drever_APB1983}
R.~W.~P. Drever, J.~L. Hall, F.~V. Kowalski, J.~Hough, G.~M. Ford, A.~J.
  Munley, and H.~Ward.
\newblock Laser phase and frequency stabilization using an optical resonator.
\newblock 31(2):97--105, 1983.

\bibitem{Li_OptLett2006}
Chengquan Li, Eric Moon, and Zenghu Chang.
\newblock Carrier-envelope phase shift caused by variation of grating
  separation.
\newblock {\em Opt. Lett.}, 31(21):3113--3115, Nov 2006.

\bibitem{Treacy_IEEE1969}
E.~Treacy.
\newblock Optical pulse compression with diffraction gratings.
\newblock {\em IEEE Journal of Quantum Electronics}, 5(9):454--458, 1969.

\bibitem{Wise_PRL2005}
Stacy Wise, V.~Quetschke, A.~J. Deshpande, G.~Mueller, D.~H. Reitze, D.~B.
  Tanner, B.~F. Whiting, Y.~Chen, A.~T\"unnermann, E.~Kley, and T.~Clausnitzer.
\newblock Phase effects in the diffraction of light: Beyond the grating
  equation.
\newblock {\em Phys. Rev. Lett.}, 95:013901, 2005.

\bibitem{Zhou_OptExp2005}
Shian Zhou, Lyuba Kuznetsova, Andy Chong, and Frank~W. Wise.
\newblock Compensation of nonlinear phase shifts with third-order dispersion in
  short-pulse fiber amplifiers.
\newblock {\em Opt. Express}, 13(13):4869--4877, Jun 2005.

\bibitem{Roser_OptLett2005}
F.~R\"{o}ser, J.~Rothhard, B.~Ortac, A.~Liem, O.~Schmidt, T.~Schreiber,
  J.~Limpert, and A.~T\"{u}nnermann.
\newblock 131 w 220 fs fiber laser system.
\newblock {\em Opt. Lett.}, 30(20):2754--2756, Oct 2005.

\bibitem{Zhao_AppPhysExp2016}
Zhigang Zhao and Yohei Kobayashi.
\newblock Ytterbium fiber-based, 270 fs, 100 w chirped pulse amplification
  laser system with 1 mhz repetition rate.
\newblock {\em Applied Physics Express}, 9(1):012701, 2016.

\bibitem{Wunram_OptLett2015}
Marcel Wunram, Patrick Storz, Daniele Brida, and Alfred Leitenstorfer.
\newblock Ultrastable fiber amplifier delivering 145-fs pulses with 6-$\mu$j
  energy at 10-mhz repetition rate.
\newblock {\em Opt. Lett.}, 40(5):823--826, Mar 2015.

\bibitem{Ruehl_OPN2012}
Axel Ruehl.
\newblock Advances in yb:fiber frequency comb technology.
\newblock {\em Optics and Photonics News}, May 2012:31, 2012.

\bibitem{Mills_SPIE2015}
Sheyerman A. Levy G. Damascelli~A. Mills A.~K., Zhadnovich~S. and Jones~D. J.
\newblock An xuv source using a femtosecond enhancement cavity for
  photoemission spectroscopy.
\newblock {\em SPIE}, 2015.

\bibitem{Wu_FiO2013}
Tsung-Han Wu, David Carlson, and R.Jason Jones.
\newblock A high-power fiber laser system for dual-comb spectroscopy in the
  vacuum-ultraviolet.
\newblock In {\em Frontiers in Optics 2013}, page FTu2A.4. Optical Society of
  America, 2013.

\bibitem{Offner_Patent}
A.~Offner and D.~Conn.
\newblock U. S. Patent Number 3748015, 1973.

\bibitem{Cheriaux_OptLett1996}
G.~Cheriaux, Barry Walker, L.~F. Dimauro, P.~Rousseau, F.~Salin, and J.~P.
  Chambaret.
\newblock Aberration-free stretcher design for ultrashort-pulse amplification.
\newblock {\em Opt. Lett.}, 21(6):414--416, Mar 1996.

\bibitem{Kane_JOSAB1997}
S.~Kane and J.~Squier.
\newblock Fourth-order-dispersion limitations of aberration-free chirped-pulse
  amplification systems.
\newblock {\em J. Opt. Soc. Am. B}, 14(5):1237--1244, 1997.

\bibitem{Kuznetsova_OSA2007}
Lyuba Kuznetsova, Frank~W. Wise, Steve Kane, and Jeff Squier.
\newblock Chirped-pulse amplification near the gain-narrowing limit of an
  yb-doped fiber amplifier using a reflection grism compressor.
\newblock In {\em Conference on Lasers and Electro-Optics/Quantum Electronics
  and Laser Science Conference and Photonic Applications Systems Technologies},
  page CMEE7. Optical Society of America, 2007.

\bibitem{Siegman}
Anthony~E. Siegman.
\newblock {\em Lasers}.
\newblock University Science Books, 1986.

\bibitem{Gherman_OptExp2002}
Titus Gherman and Daniele Romanini.
\newblock Modelocked cavity--enhanced absorption spectroscopy.
\newblock {\em Opt. Express}, 10(19):1033--1042, Sep 2002.

\bibitem{Corder_DAMOP2016}
C.~Corder, P.~Zhao, X.~L. Li, A.~R. Muraca, M.~D. Kershis, M.~G. White, and
  T.~K. Allison.
\newblock Ultrafast xuv pulses at high repetition rate for time resolved
  photoelectron spectroscopy of surface dynamics.
\newblock In {\em Bulletin of the Americal Physical Society, 47th Annual
  Meeting of the APS Division of Atomic, Molecular and Optical Physics
  (DAMOP)}, 2016.

\bibitem{Davis_ChemPhysLett2001}
Scott Davis, Michal Farnik, Dairene Uy, and David~J. Nesbitt.
\newblock Concentration modulation spectroscopy with a pulsed slit supersonic
  discharge expansion source.
\newblock {\em Chemical Physics Letters}, 344(1‚{\"A}{\`\i}2):23 -- 30, 2001.

\bibitem{Ye_JOSAB1998}
Jun Ye, Long-Sheng Ma, and John~L. Hall.
\newblock Ultrasensitive detections in atomic and molecular physics:
  demonstration in molecular overtone spectroscopy.
\newblock {\em J. Opt. Soc. Am. B}, 15(1):6--15, Jan 1998.

\bibitem{Gagliardi_Book2013}
Gianluca Gagliardi and Hans-Peter Loock, editors.
\newblock {\em Cavity Enhanced Spectroscopy and Sensing}.
\newblock Springer, 2013.

\bibitem{Wang_JPhysChem1995}
Juen-Kai Wang, Qianli Liu, and Ahmed~H. Zewail.
\newblock Solvation ultrafast dynamics of reactions. 9. femtosecond studies of
  dissociation and recombination of iodine in argon clusters.
\newblock {\em The Journal of Physical Chemistry}, 99(29):11309--11320, 1995.

\bibitem{Brabec_RMP2000}
T.~Brabec and F.~Krausz.
\newblock Intense few cycle laser fields: Frontiers of nonlinear optics.
\newblock {\em Rev. Mod. Phys.}, 72:545--591, 2000.

\bibitem{Chini_nph2014}
Michael Chini, Kun Zhao, and Zenghu Chang.
\newblock The generation, characterization and applications of broadband
  isolated attosecond pulses.
\newblock {\em Nat Photon}, 8(3):178--186, 03 2014.

\bibitem{Passlack_JAP2006}
S.~Passlack, S.~Mathias, O.~Andreyev, D.~Mittnacht, M.~Aeschlimann, and
  M.~Bauer.
\newblock Space charge effects in photoemission with a low repetition, high
  intensity femtosecond laser source.
\newblock {\em Journal of Applied Physics}, 100(2), 2006.

\bibitem{Sandhu_Science2008}
Arvinder~S. Sandhu, Etienne Gagnon, Robin Santra, Vandana Sharma, Wen Li, Phay
  Ho, Predrag Ranitovic, C.~Lewis Cocke, Margaret~M. Murnane, and Henry~C.
  Kapteyn.
\newblock {Observing the Creation of Electronic Feshbach Resonances in Soft
  X-ray-Induced O-2 Dissociation}.
\newblock {\em Science}, {322}({5904}):{1081--1085}, {2008}.

\bibitem{Gohle_Nature2005}
C.~Gohle, T.~Udem, M.~Herrmann, J.~Rauschenberger, R.~Holzwarth, H.~A.
  Schuessler, F.~Krausz, and T.~W. H\"{a}nsch.
\newblock {A frequency comb in the extreme ultraviolet}.
\newblock {\em Nature}, {436}({7048}):{234--237}, {2005}.

\bibitem{Jones_PRL2005}
R.~Jason Jones, Kevin~D. Moll, Michael~J. Thorpe, and Jun Ye.
\newblock Phase-coherent frequency combs in the vacuum ultraviolet via
  high-harmonic generation inside a femtosecond enhancement cavity.
\newblock {\em Physical Review Letters}, 94(19):193201, 2005.

\bibitem{Allison_PRL2011}
T.~K. Allison, A.~Cing\"oz, D.~C. Yost, and J.~Ye.
\newblock Extreme nonlinear optics in a femtosecond enhancement cavity.
\newblock {\em Phys. Rev. Lett.}, 107:183903, Oct 2011.

\bibitem{Carlson_OptLett2011}
D.~R. Carlson, Jane Lee, John Mongelli, E.~M. Wright, and R.~J. Jones.
\newblock Intracavity ionization and pulse formation in femtosecond enhancement
  cavities.
\newblock {\em Opt. Lett.}, 36(15):2991--2993, Aug 2011.

\bibitem{Yost_OptExp2011}
D.~C. Yost, A.~Cing\"{o}z, T.~K. Allison, A.~Ruehl, M.~E. Fermann, I.~Hartl,
  and J.~Ye.
\newblock Power optimization of xuv frequency combs for spectroscopy
  applications [invited].
\newblock {\em Opt. Express}, 19(23):23483--23493, Nov 2011.

\bibitem{Holzberger_PRL2015}
S.~Holzberger, N.~Lilienfein, H.~Carstens, T.~Saule, M.~H\"ogner, F.~L\"ucking,
  M.~Trubetskov, V.~Pervak, T.~Eidam, J.~Limpert, A.~T\"unnermann, E.~Fill,
  F.~Krausz, and I.~Pupeza.
\newblock Femtosecond enhancement cavities in the nonlinear regime.
\newblock {\em Phys. Rev. Lett.}, 115:023902, Jul 2015.

\bibitem{Pupeza_PRL2014}
I.~Pupeza, M.~H\"ogner, J.~Weitenberg, S.~Holzberger, D.~Esser, T.~Eidam,
  J.~Limpert, A.~T\"unnermann, E.~Fill, and S.~Yakovlev, V.\.
\newblock Cavity-enhanced high-harmonic generation with spatially tailored
  driving fields.
\newblock {\em Phys. Rev. Lett.}, 112:103902, Mar 2014.

\bibitem{Lee_OptExp2011}
Jane Lee, David~R. Carlson, and R.~Jason Jones.
\newblock Optimizing intracavity high harmonic generation for xuv fs frequency
  combs.
\newblock {\em Opt. Express}, 19(23):23315--23326, Nov 2011.

\bibitem{Frassetto_OptExp2011}
Fabio Frassetto, Cephise Cacho, Chris~A. Froud, I.C.~Edmund Turcu, Paolo
  Villoresi, Will~A. Bryan, Emma Springate, and Luca Poletto.
\newblock Single-grating monochromator for extreme-ultraviolet ultrashort
  pulses.
\newblock {\em Opt. Express}, 19(20):19169--19181, Sep 2011.

\bibitem{Chew_BookChapter2015}
S.~H. Chew, K.~Pearce, C.~SPath, A.~Guggenmos, J.~Schmidt, F.~Sussman, M.~F.
  Kling, U.~Kleinberg, E.~Marsell, A.~L. Cord, E.~Lorek, P.~Rudawski, C.~Guo,
  M.~Miranda, F.~Ardana, J.~Mauritsson, A.~L'Huillier, and A.~Mikkelsen.
\newblock Imaging localized surface plasmons by femtosecond to attosecond
  time-resolved photoelectron emission microscopy.
\newblock In {\em Attosecond Nanophysics}. WILEY-VCH Verlag, 2015.

\bibitem{Plotzing_RSI2016}
M.~Plotzing, R.~Adam, C.~Weier, L.~Plucinski, S.~Eich, S.~Emmerich,
  M.~Rollinger, M.~Aeschlimann, S.~Mathias, and C.~M. Schneider.
\newblock Spin-resolved photoelectron spectroscopy using femtosecond extreme
  ultraviolet light pulses from high-order harmonic generation.
\newblock {\em Review of Scientific Instruments}, 87(4), 2016.

\end{thebibliography}

\clearpage

\begin{widetext}
\begin{center}

\large{\textbf{Supplementary information for high-power ultrafast Yb:fiber laser frequency combs using commercially available components and basic fiber tools}}

\small{X. L. Li, M. A. R. Reber, C. Corder, Y. Chen, P. Zhao, and T. K. Allison}

Stony Brook University, Stony Brook, NY 11794-3400 USA

\end{center}


\section*{\normalsize{\leftline{S1. INFORMATION OF PARTS}}}

The tables in this section list the detailed information for the parts used in our laser system (oscillators and amplifiers). 

\begin{table}[ht]
\centering  
\begin{tabular}{l|l|l|l}  
\hline
\hline
Part &Description  &Oscillator A &Oscillator B \\ \hline
\hline
M1&D-shaped mirror &\multicolumn{2}{l}{PFD10-03-P01}\\ \hline
M2 &Protected silver mirror &\multicolumn{2}{l}{PF10-03-P01}\\ \hline
RP &Roof reflecting prism &\multicolumn{2}{l}{PS908H-C}\\ \hline
HWP &Zeroth-order half-wave plate &WPH05M-1053 &WPH05M-1030\\ \hline
QWP1 \& &Zeroth-order quarter-wave plate &WPQ05M-1053 &WPQ05M-1030\\ 
QWP2        &                                                                &                            &                             \\\hline
PBS &Polarizing beam splitter &PBS123 &PBS103\\ \hline
G1 \& &Grating &Lightsmyth              &Wasatch Photonics\\ 
G2        &               &1000 groove/mm &600 groove/mm\\ \hline
EOM &Electro-optical modulator&EO-PM-NR-C2 &United Crystals, LiTaO3, Y-cut \\ 
          &                                                  &                           &AR-coated on 2 ends, 4 mm cube  \\ \hline
WDM &Wavelength-division multiplexing&WD202G &WD202F\\ \hline
Yb:Fiber& &YB1200-6/125DC, 26 cm &YB1200-4/125, 20 cm\\ \hline
SM fiber& Single mode fiber &P4-980AR-2 \& WDM pigtails &P4-980AR-2 \& WDM pigtails\\ 
               &                                    &OFS980, totally 164 cm &OFS980, totally 170 cm\\ \hline
\multicolumn{2}{l|}{976 nm Pump Laser} &\multicolumn{2}{l}{Oclaro Technology Limited LC96L76P-20R}\\ \hline
\multicolumn{2}{l|}{Fiber port collimator} &PAF-X-5-C &PAF-X-5-B\\ \hline
I1 &Isolator  &IO-3D-1053-VLP &IO-3D-1030-VLP\\ \hline
I2 &Isolator  &\multicolumn{2}{l}{Isolator, DPM Photonics HPII-98-C2-P-22-LL-1}\\ \hline
\hline
\end{tabular}
\caption{List of parts  in the oscillators (figure 2 in the paper). Part numbers are from Thorlabs unless otherwise indicated.}
\label{table1}
\end{table}
\begin{table}[ht]
\centering  
\begin{tabular}{l|l}  
\hline
\hline
Part Name  &Details \\ \hline
Stretcher  &OFS Fitel Denmark ApS, GDD=1.65 fs$^2$,TOD=-7.9$\times$10$^6$ fs$^3$\\\hline  
PC  &Polarization Controller, PLC-900\\\hline  
L1  &Lens, A110-TM-B, f = 6.24 mm\\\hline  
L2  &Lens, LA1289-B, f = 30 mm\\\hline     
L3  &Lens, AL4532-B, f = 32 mm\\\hline
L4  &Lens, AL3026-B, f = 26 mm\\\hline
L5  &Lens, LC4210-C, f =-25 mm\\\hline
L6  &Lens, LA4725-C, f = 75 mm\\\hline
D1 &Layertec 103675 short wave pass filter\\\hline
D2 &Layertec 108834 long wave pass filter\\\hline
I1  &Isolator, IO-3D-1053-VLP\\\hline
I2 \& I3 &Isolator, EOT 1200462\\\hline
HWP  &Zeroth-order half-wave plate, WPH10M-1053\\\hline
M1  &Altechna low GDD HR broadband mirror \\\hline
M2  &BBSQ1-E03 square broadband mirror \\\hline
5 m PCF &NKT Photonics DC-200/40-PZ-Yb Aerogain flex 5.0, MFD = 31 $\mu$m  \\\hline
915 nm pump laser & nLight  Photonics, element-E3-1040677, maximum output 30 W \\\hline
915 nm laser driver &Vuemetrix, Vue-MV12-0112V-10A \\\hline 
G1 \& G2 &1250 groove/mm, Wasatch Photonics custom diffraction gratings\\ \hline
RM & Two mirrors, Altechna low GDD HR broad band mirror, size: $3''\times 1''$\\ \hline
PD  &Photodiode, PDA10CD\\\hline
\hline
\end{tabular}
\caption{List of parts in the 9 W optical amplifier (figure 6 in the paper). Part numbers are from Thorlabs unless otherwise indicated. GDD = group delay dispersion. TOD = third-order dispersion. f = focal length. MFD = mode-field diameter.}
\label{table2}
\end{table}
\begin{table}[ht]
\centering  
\begin{tabular}{l|l}  
\hline
\hline
Part Name  &Details \\ \hline
Stretcher  &OFS Fitel Denmark ApS, GDD=5.27 fs$^2$,TOD=-2.16$\times$10$^7$ fs$^3$\\\hline  
PC  &Polarization Controller, PLC-900\\\hline  
L1  &Lens, A110TM-B, f = 6.24 mm\\\hline  
L2  &Lens,  LA1289-B, f = 30 mm\\\hline     
L3  &Lens,   AL4532-B, f = 32 mm\\\hline
L4  &Lens,   AL3026-B, f = 26 mm\\\hline
L5  &Lens,   LA1134-B, f = 60 mm\\\hline
L6  &Lens,   AL4532-B, f = 32 mm\\\hline
L7  &Lens,   AL5040-B, f = 40 mm\\\hline
L8  &Lens,   LC4252-B, f =-30 mm\\\hline 
L9  &Lens,   LA4874-B, f = 150 mm\\\hline 
D1 &Layertec 108834 long wave pass filter\\\hline
D2 &Layertec 103675 short wave pass filter\\\hline
D3 &Layertec 109842 short wave pass filter\\\hline
I1  &Isolator, IO-3D-1030-VLP\\\hline
I2 \& I3 &Isolator, EOT 1200462\\\hline
HWP  &Zeroth-order half-wave plate, WPH10M-1030\\\hline
M1  &Altechna low GDD HR broadband mirror \\\hline
M2  &BB1-E03 broadband mirror \\\hline
M3  &BBSQ1-E03 square broadband mirror \\\hline
2.5 m PCF &NKT Photonics DC-200/40-PZ-Yb Aerogain flex 2.5, MFD = 31 $\mu$m  \\\hline
915 nm pump laser & nLight  Photonics, element-E3-1040677, maximum output 30 W \\\hline
915 nm laser driver &Vuemetrix, Vue-MV12-0112V-10A \\\hline 
0.8 m PCF Rod &NKT Photonics aeroGAIN-ROD-PM85, MFD = 65 $\mu$m \\\hline
975 nm pump laser &LIMO GmbH, LIMO200-F200-DL980-S1886, maximum output 200 W \\\hline 
975 nm laser driver &Lambda Inc, Genesys 1U Series 1500W Power Supply GEN 20-76\\\hline 
BD1 \& BD2 &Beam dump \\\hline  
G1 \& G2 &1250 groove/mm, Wasatch Photonics custom diffraction gratings\\ \hline
RM &Two mirrors, Altechna low GDD HR broadband mirror, size: $3''\times 1''$\\ \hline
PD  &Photodiode, PDA10CF\\\hline
\hline
\end{tabular}
\caption{List of parts in the 80 W optical amplifier (figure 8 in the paper.) Part numbers are from Thorlabs unless otherwise indicated.}
\label{table3}
\end{table}

\clearpage

\section*{\normalsize{\leftline{S2. BEAM DUMP}}}

\begin{figure}[ht]
   \centering
   \includegraphics[width=8cm]{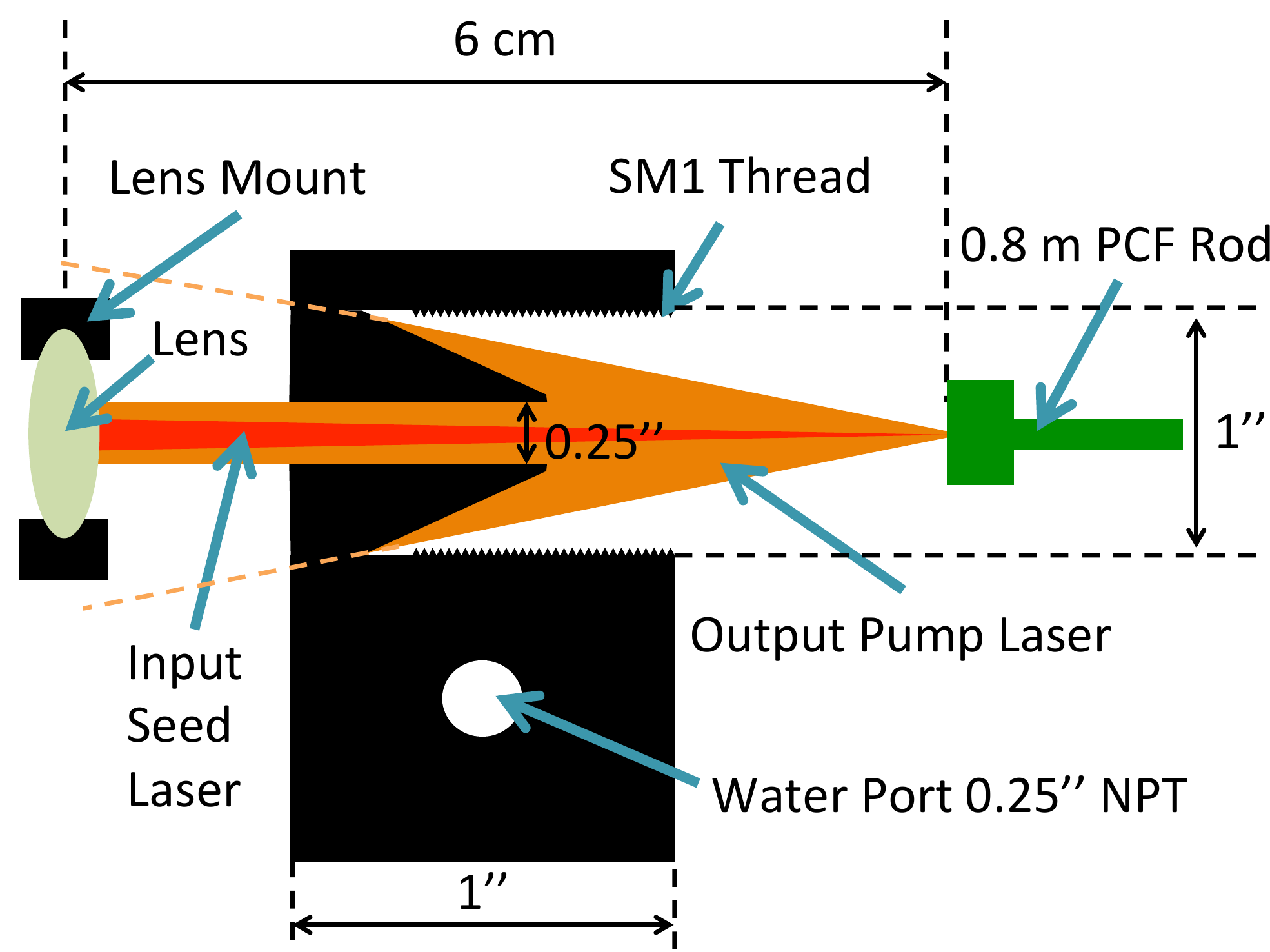} 

\begin{center}
Figure S1. Detailed description of BD2  from figure 8 in the paper.
\end{center}
   \label{BD}
\end{figure}

Figure S1 shows how the beam dump (BD2 in the 80 W laser system) works to protect the lens mount from exposure to the high power pump light. Due to the large numerical aperture (NA=0.5) of the inner cladding of the 0.8 m photonic-crystal fiber (PCF) rod, the powerful  pump laser coming out of the fiber has a very big divergence angle and will heat the mount of the lens if the water-cooled beam dump is absent. The beam dump blocks most of the pump light but allows the seed light to go through.

\section*{\normalsize{\leftline{S3. WAVELENGTH OF PUMP LASER IN THE 80 W AMPLIFIER}}}

\begin{figure}[htbp]
   \centering
   \includegraphics[width=8cm]{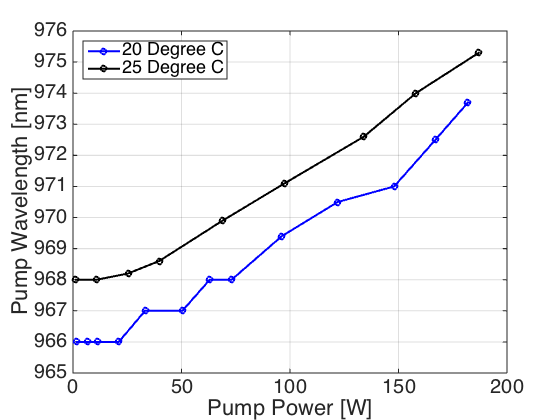} 
   \begin{center}
Figure S2. The output wavelength of 975 nm pump laser at different cooling water temperatures and and different power.
\end{center}
   \label{Pwavelength}
\end{figure}

In the 80 W amplifier, as shown in figure S2, the output wavelength of the 975 nm pump diode changes with the operation temperature and power in the range of 967 nm to 977 nm, causing an increasing amplification efficiency while turning up the pump power (figure 9a) in the paper). The temperature shown is just for the cooling water flowing into the pump diode. The pump diode temperature is not controlled.

\section*{\normalsize{\leftline{S4. MECHANICAL DESIGN OF AMPLIFIERS}}}

\begin{figure}[htbp]
   \centering
   \includegraphics[width=10cm]{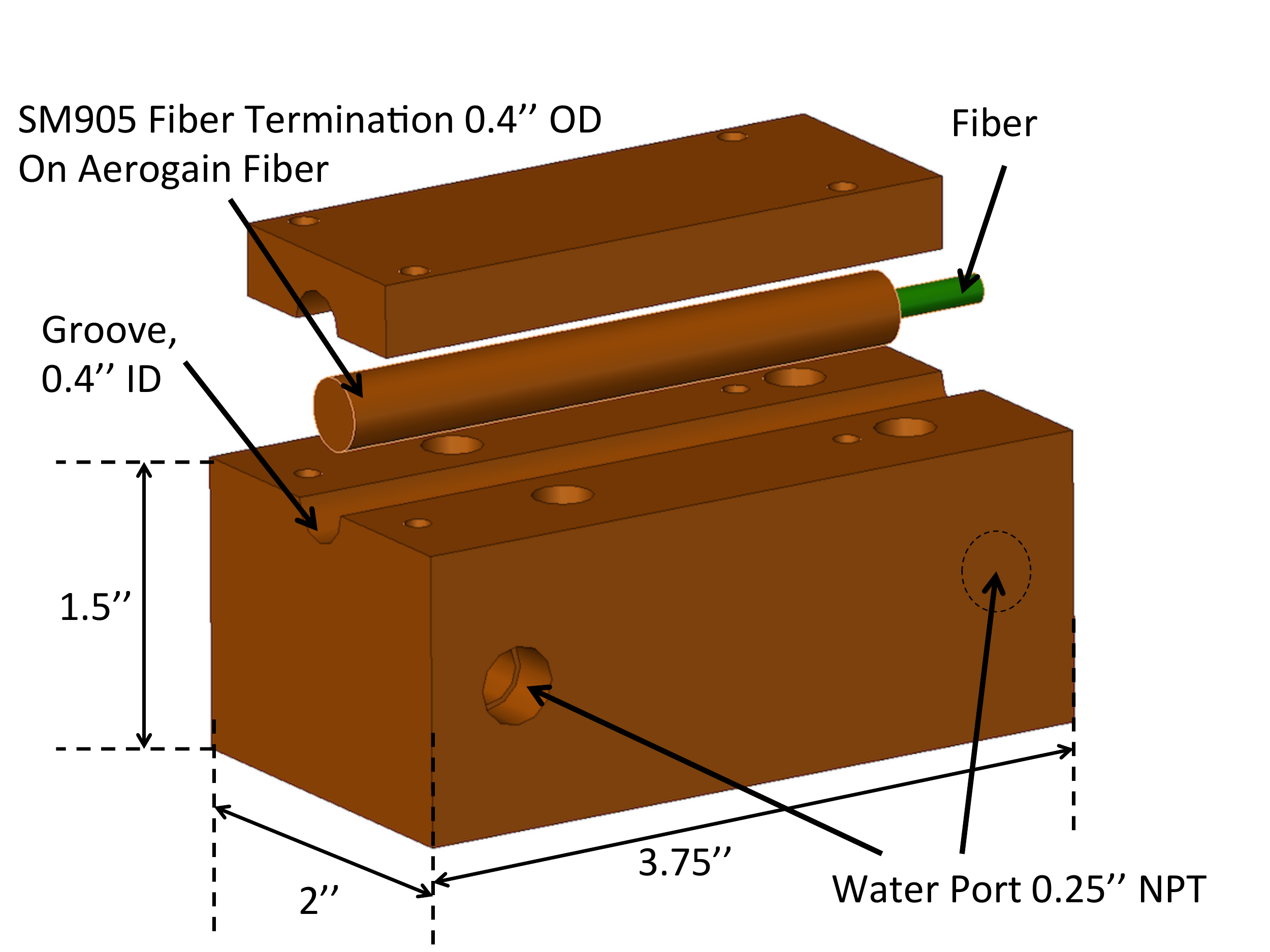} 
   \begin{center}
Figure S3. The cooling solution for the pump end of the coiled NKT LMA double-clad PCF.
\end{center}
   \label{Pwavelength}
\end{figure}

Figure S3 shows the pump end of the coiled  large mode area (LMA) double-clad PCF  (5 m in the 9 W laser and 2.5 m in the 80 W laser). The pump power and the seed power are both low at the seed end, so we only  cool the pump end of the fiber (25 W pump laser and 12 W seed laser is guided in the pump end). The pump end, together with copper SMA905 termination (NKT Photonics) is placed in a water-cooled copper clamshell. The termination is wrapped in a piece of indium foil to increase the thermal conduction with the copper clamshell.

\begin{figure}[htbp]
   \centering
   \includegraphics[width=10cm]{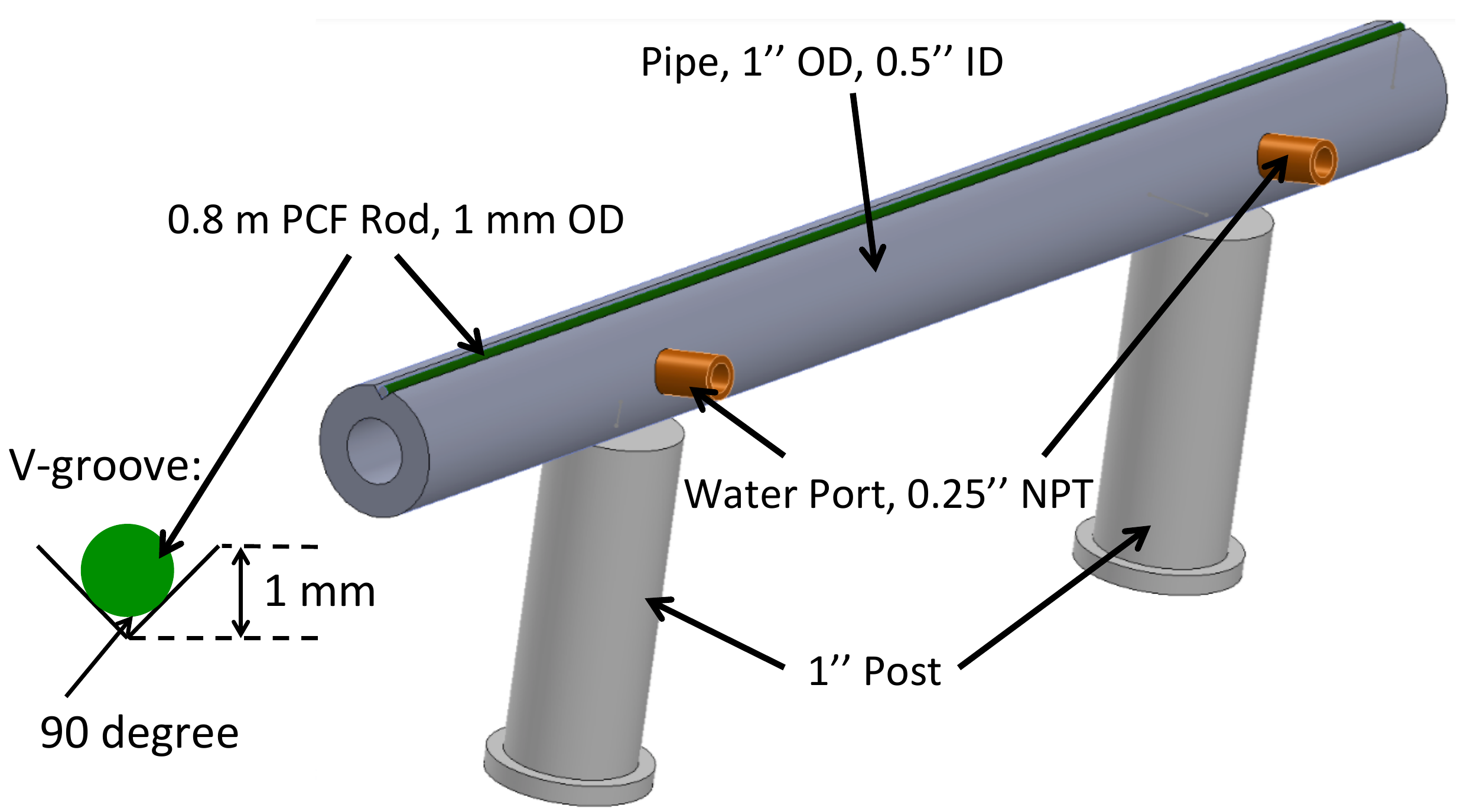} 
   \begin{center}
Figure S4. Cross section of the cooling solution for NKT 0.8 m PCF rod.
\end{center}
   \label{Pwavelength}
\end{figure}

Figure S4 shows the cooling solution for the NKT 0.8 m PCF rod.  The rod fiber, whose diameter is 1 mm, is placed in the V-groove of 1 inch aluminum pipe, held loosely by only two pieces of kapton tape,  and water flows between the water ports. The ends of the pipe (0.5 inch inner diameter) are plugged with NPT fittings.

As seen in figure S5, the end cap of the rod fiber has a dimension of 6 mm (OD) $\times$5 mm (length). So it can not be placed in the V-groove of the aluminum pipe.  A separation V-jaw supports the end cap, as shown in figure 5S.

\begin{figure}[t]
   \centering
   \includegraphics[width=8cm]{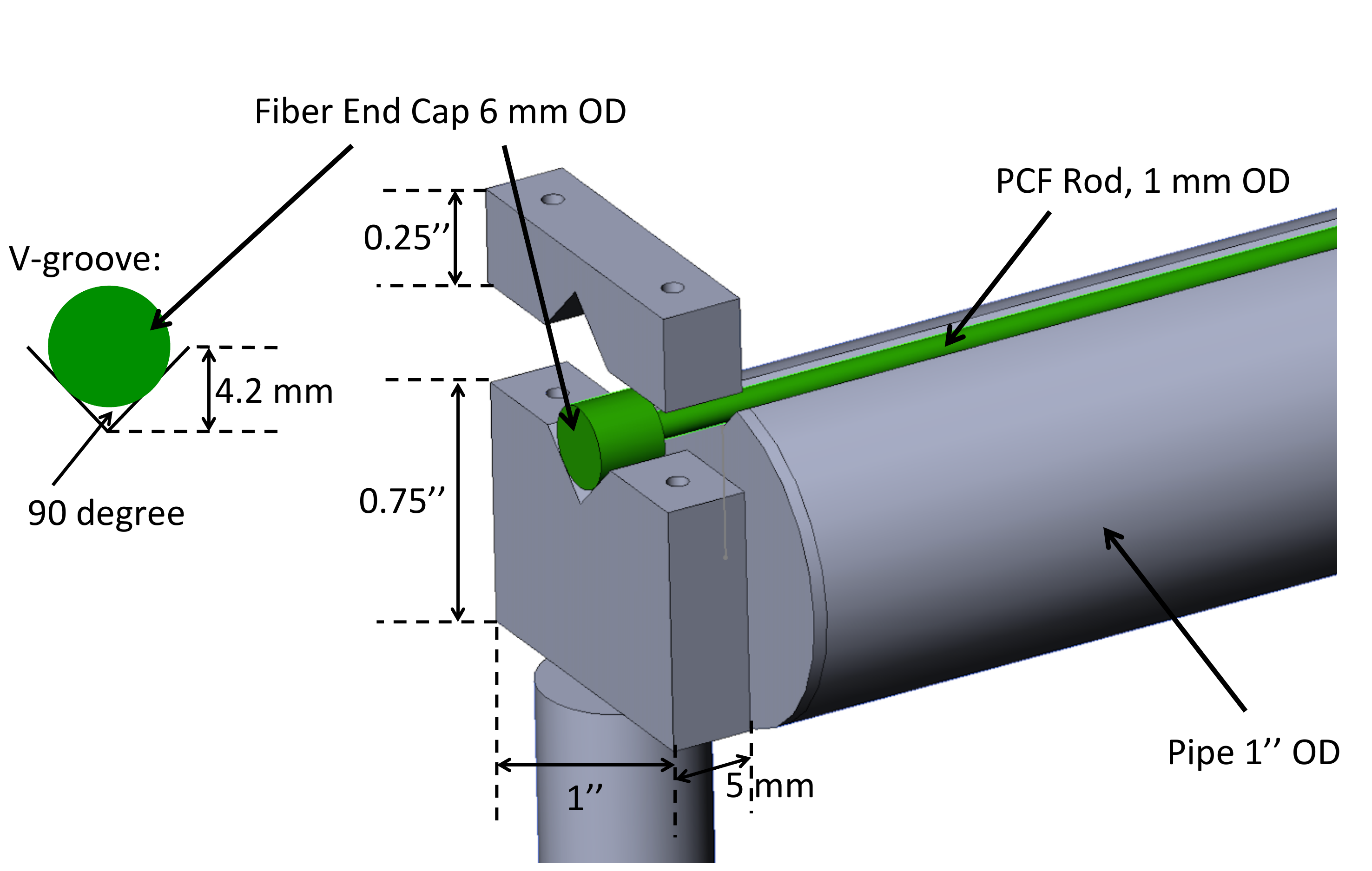} 
   \begin{center}
Figure S5. The design of the support for the end cap of the 0.8 m NKT PCF rod.
\end{center}
   \label{Pwavelength}
\end{figure}

\end{widetext}

\end{document}